\documentclass[journal, compsoc]{IEEEtran}
\usepackage[nocompress]{cite}
\usepackage[hidelinks]{hyperref}
\usepackage{url}
\usepackage{bookmark}
\usepackage{alltt}
\usepackage{multirow}
\usepackage{graphicx}
\usepackage{color}
\usepackage{pifont}
\usepackage{boxedminipage}
\usepackage{notoccite}
\usepackage{array}
\newcolumntype{C}[1]{>{\centering\let\newline\\\arraybackslash\hspace{0pt}}m{#1}}
\usepackage{enumerate}
\usepackage{booktabs}
\usepackage[framemethod=tikz]{mdframed}
\usepackage{lipsum}
\usepackage{mathtools}
\usepackage{paralist}
\usepackage{boldline}
\usepackage[utf8x]{inputenc}
\usepackage{colortbl} 
\usepackage{blindtext, graphicx}
\usepackage{textcomp}
\usepackage{mathtools}
\usepackage{amsmath}
\usepackage{amsfonts}
\usepackage{color}
\usepackage[final]{listings}
\usepackage{graphicx} 
\usepackage{multirow}
\usepackage{balance}
\usepackage{picture}
\usepackage{relsize}
\usepackage{multicol}
\usepackage{soul}
\usepackage{enumitem}
\usepackage{array}
\usepackage{makecell}
\usepackage{bigstrut}
\usepackage{array}
\usepackage{makecell}
\usepackage[skins]{tcolorbox}
\usepackage{verbatim}
\usepackage{algorithmicx}
\usepackage{algpseudocode}
\usepackage[export]{adjustbox}
\usepackage{mathtools}
\usepackage{fancyvrb}
\usepackage{soul}
\usepackage{color}
\usepackage{fancybox}
\usepackage{multirow}
\usepackage{listings}
\usepackage{mathtools}
\usepackage{url}
\usepackage{float}
\usepackage{comment}
\usepackage{framed}
\usepackage{balance}
\usepackage[turnon]{notes}
\usepackage{xspace}
\usepackage{slashbox}
\usepackage{csquotes}
\usepackage[T1]{fontenc}
\usepackage{enumitem}
\usepackage{booktabs} 
\usepackage[english]{babel}
\usepackage{amsthm}
\usepackage[linesnumbered, ruled]{algorithm2e}
\usepackage[noabbrev]{cleveref}
\usepackage{etoolbox}
\newcommand{\subparagraph}{}
\newcommand{\tion}[1]{\S\ref{sect:#1}}
\newcommand{\fig}[1]{Figure~\ref{fig:#1}}
\newcommand{\tab}[1]{Table~\ref{tab:#1}}
\definecolor{comment_color}{rgb}{0.5, 0, 1}
\definecolor{result}{rgb}{0.1, 0.3, 0.5} 
\definecolor{steel}{rgb}{0, 0.2, 0.9} 

\newcommand{\bi}{\begin{itemize}}
\newcommand{\be}{\begin{enumerate}}
\newcommand{\ei}{\end{itemize}}
\newcommand{\ee}{\end{enumerate}}
\newcommand{\eq}[1]{Equation~\ref{eq:#1}}

\makeatletter
\newcommand{\removelatexerror}{\let\@latex@error\@gobble}
\makeatother
\sethlcolor{lightgray}
\definecolor{pink}{RGB}{252,145,149}
\definecolor{lightpink}{RGB}{252,145,149}
\definecolor{lightgray}{gray}{0.8}
\definecolor{darkgray}{gray}{0.6}
\definecolor{Gray}{rgb}{0.88,1,1}
\definecolor{Gray}{gray}{0.85}
\definecolor{Blue}{RGB}{0,29,193}
\definecolor{MyDarkBlue}{rgb}{0,0.08,0.45} 
\definecolor{pink}{RGB}{231,95,110}
\definecolor{greenish}{RGB}{182, 231, 142}
\definecolor{resultbox}{HTML}{0D3d56}
\definecolor{orangish}{RGB}{255, 206, 144}
\definecolor{lavender}{RGB}{225, 213, 231}
\definecolor{lightergray}{rgb}{0.85, 0.85, 0.85}
\definecolor{lightestgray}{rgb}{0.95, 0.95, 0.95}
\definecolor{codebg}{HTML}{F4F4F4}
\definecolor{blueish}{RGB}{177, 206, 232}
\definecolor{resultblue}{RGB}{199, 210, 224}
\definecolor{fig1_blue}{RGB}{85, 85, 255}
\definecolor{fig1_green}{RGB}{34, 148, 34}
\definecolor{gray05}{gray}{0.95}
\definecolor{gray08}{gray}{0.92}
\definecolor{gray10}{gray}{0.90}
\definecolor{gray12}{gray}{0.88}
\definecolor{gray15}{gray}{0.85}
\definecolor{gray18}{gray}{0.82}
\definecolor{gray20}{gray}{0.80}
\definecolor{gray25}{gray}{0.75}
\definecolor{gray30}{gray}{0.70}
\definecolor{gray35}{gray}{0.65}
\definecolor{gray40}{gray}{0.60}
\definecolor{gray45}{gray}{0.55}
\definecolor{gray50}{gray}{0.50}
\definecolor{gray55}{gray}{0.45}
\definecolor{gray60}{gray}{0.40}
\definecolor{gray65}{gray}{0.35}
\definecolor{gray70}{gray}{0.30}
\definecolor{gray75}{gray}{0.25}
\definecolor{gray80}{gray}{0.20}
\definecolor{gray85}{gray}{0.15}
\definecolor{gray90}{gray}{0.10}
\definecolor{gray95}{gray}{0.05}

\newcommand\blue[1]{\textcolor[rgb]{0.00,0.00,1.00}{{#1}}}

\newcommand{\squishlist}{
 \begin{list}{$\bullet$}
 { \setlength{\itemsep}{0pt}
   \setlength{\parsep}{1pt}
   \setlength{\topsep}{1pt}
   \setlength{\partopsep}{0pt}
   \setlength{\leftmargin}{1em}
   \setlength{\labelwidth}{1em}
   \setlength{\labelsep}{0.5em} } }

\newcommand{\squishlisttwo}{
 \begin{list}{$\bullet$}
 { \setlength{\itemsep}{0pt}
  \setlength{\parsep}{0pt}
  \setlength{\topsep}{0pt}
  \setlength{\partopsep}{0pt}
  \setlength{\leftmargin}{0em}
  \setlength{\labelwidth}{1em}
  \setlength{\labelsep}{0.5em} } }

\newcommand{\squishend}{
 \end{list} }

\newcommand{\blind}[1]{{\em blinded}}
\newcommand{\rom}[1]{\uppercase\expandafter{\romannumeral #1\relax}}

\newcommand{\etal}{\hbox{\emph{et al.}}\xspace}
\newcommand{\eg}{\hbox{\emph{e.g.}}\xspace}
\newcommand{\ie}{\hbox{\emph{i.e.}}\xspace}
\newcommand{\wrt}{\hbox{\emph{w.r.t.}}\xspace}

\newcommand{\menquote}[1]{\ensuremath{\text{\textquotedbl} #1 \text{\textquotedbl}}}

\makeatletter
\newcommand\footnoteref[1]{\protected@xdef\@thefnmark{\ref{#1}}\@footnotemark}
\makeatother

\definecolor{gray50}{gray}{.5}
\definecolor{gray40}{gray}{.6}
\definecolor{gray30}{gray}{.7}
\definecolor{gray20}{gray}{.8}
\definecolor{gray10}{gray}{.9}
\definecolor{gray05}{gray}{.95}

\newlength\Linewidth
\def\findlength{\setlength\Linewidth\linewidth
\addtolength\Linewidth{-4\fboxrule}
\addtolength\Linewidth{-3\fboxsep}
}
\newenvironment{examplebox}{\par\begingroup
   \setlength{\fboxsep}{5pt}\findlength
   \setbox0=\vbox\bgroup\noindent
   \hsize=0.95\linewidth
   \begin{minipage}{0.95\linewidth}\normalsize}
    {\end{minipage}\egroup
    \textcolor{gray20}{\fboxsep1.5pt\fbox
     {\fboxsep5pt\colorbox{gray05}{\normalcolor\box0}}}
    \endgroup\par\noindent
    \normalcolor\ignorespacesafterend}

\newcounter{RQCounter}

\definecolor{javared}{rgb}{0.6,0,0} 
\definecolor{javagreen}{rgb}{0.25,0.5,0.35} 
\definecolor{javapurple}{rgb}{0.5,0,0.35} 
\definecolor{javadocblue}{rgb}{0.25,0.35,0.75} 

\lstdefinestyle{customc}{
  belowcaptionskip=\baselineskip,
  breaklines=true,
  xleftmargin=\parindent,
  language=java,
  showstringspaces=false,
  basicstyle=\scriptsize\ttfamily,
  keywordstyle=\bfseries\color{javapurple},
  commentstyle=\itshape\blue,
  belowskip=-10pt,
  aboveskip=-5pt
}

\newcolumntype{P}[1]{>{\centering\arraybackslash}p{#1}}

\newcommand{\tool}{\textsc{ConEx}\xspace}

\newcommand{\rqa}{Can \tool 
find configurations that are better than the baseline default configurations?\xspace}

\newcommand{\rqd}{How does the EMCMC sampling strategy perform compared to random and evolutionary sampling?\xspace}

\newcommand{\rqe}{How does \tool perform compared to the state-of-the-art Machine Learning based configuration optimization approaches?\xspace}

\newcommand{\respto}[1]{
\fcolorbox{black}{black!15}{%
\label{resp:#1}%
\bf\scriptsize{\color{black}R{#1}}}\xspace%
}

\xspace%

\usepackage[framemethod=tikz]{mdframed}
\usetikzlibrary{shadows}
\usepackage{graphics}
\newmdenv[
    tikzsetting= {fill=steel!08},
    skipabove=0.33em,
    skipbelow=0.33em,
    linewidth=1.33pt,
    innerleftmargin=3pt,
    innerrightmargin=3pt,
    innertopmargin=3pt,
    innerbottommargin=2pt,
    linecolor=black,
    roundcorner=2pt, 
    shadow=true,
    shadowsize=4.5pt,
    shadowcolor=black
]{myshadowbox}

\newcommand*\circled[1]{\tikz[baseline=(char.base)]{
            \node[shape=circle,draw,inner sep=1pt,fill=black!1,line width=0.1pt] (char) {{\color{black}#1}};}}

\newenvironment{result}
{\vspace{0.15cm}
\noindent\begin{minipage}{\linewidth}
\begin{center}
\arrayrulecolor{black}
\color{black}
\begin{tabular}{|p{0.95\linewidth}|}
\hline%
\rowcolor{steel!08}\textbf{Result:}%
}
{
\\\hline
\end{tabular}
\end{center}
\end{minipage}
\vspace{0.15cm}
}

\newcommand{\algoevolve}{%
\begingroup
\removelatexerror
\begin{algorithm*}[H]
    \small
  \caption{{The evolutionary sub-routine of EMCMC}}
  \label{alg:evolution}

  \SetKwInOut{Input}{Input} 
   \SetKwInOut{Output}{Output}
   \SetKwRepeat{Do}{do}{while}
   
 \underline{Function Evolve}   \\
  \Input{$conf_{best}, confs_{accepted}$} 
  \Output{$conf_{children}$}
    $conf_{children}$ $\leftarrow$ $EmptyList$ \label{alg:evolve:init_offspring} \\
    $P_{crossover}$ $\leftarrow$ randomly select 50\% parameters from $conf_{best}$ \label{alg:evolve:get_crossover_parameters} \\
    $P_{mutate}$ $\leftarrow$ randomly select 6\% parameters of $conf_{best}$ \label{alg:evolve:get_mutate_parameters}\\
    \ForEach{$conf_p \in confs_{accepted}$} { \label{alg:evolve:iterate_all_parents}
        $conf_{new}$ $\leftarrow$ $crossover(conf_{best}, conf_{p}, P_{crossover})$ \label{alg:evolve:crossover}\\
        $conf_{new}$ $\leftarrow$ $mutate(conf_{new}, P_{mutate})$ \label{alg:evolve:mutate}\\
        $conf_{children}.add(conf_{new})$ \label{alg:evolve:add_into_children}
    }
    \Return{$conf_{children}$} \label{alg:evolve:return}
\end{algorithm*}
\endgroup}

\newcommand{\algoemcmc}{%
\begingroup
\removelatexerror
\begin{algorithm*}[H]
    \small
    \setlength\textfloatsep{1\baselineskip plus 0pt minus 0 pt}
    \setlength\intextsep{1\baselineskip plus 0pt minus 0 pt}
      \caption{{Explore Configuration Space}}
      \label{alg:mcmc-ga}
    
      \SetKwInOut{Input}{Input} 
       \SetKwInOut{Output}{Output}
       \SetKwRepeat{Do}{do}{while}
       
     \underline{Function EMCMC()}   \\
      \Input{Refined Configuration Space $\zeta$, 
                job, \\
                seed $conf_{seed}$, 
                threshold $max\_gen$
                }   \label{alg:mcmc-ga:line:input}
      \Output{Best configuration $conf_{best}$}  \label{alg:mcmc-ga:line:ouput}
    
         $perf_{seed}$ $\leftarrow$ run $job$ with $conf_{seed}$ \label{alg:mcmc-ga:line:init_perf}\\
         $conf_{best}$, $perf_{best}$  $\leftarrow$ $conf_{seed}$, $perf_{seed}$ \\
         $generation$ $\leftarrow$ $1$ \label{alg:mcmc-ga:init_gens}\\
         $\Delta perf$ $\leftarrow$ $0$ \label{alg:mcmc-ga:init_global_imp}\\
              $conf_{parents}$ $\leftarrow$  sample $n$ random configurations from $\zeta$ \label{alg:mcmc-ga:line:sample_fst_generation} \\
         \While{generation $<$ max\_gen } { 
            $confs_{accepted}$ $\leftarrow$ $EmptyList$ \\
            \ForEach{parent $conf_p \in conf_{parents}$} { \label{alg:mcmc-ga:line:loop_start}
                    $perf_{p}$ $\leftarrow$ run $job$ with $conf_p$ configuration \label{alg:mcmc-ga:line:evaluate_confp} \\
                    $accepted$ $\leftarrow$ Accept($perf_{best}, perf_{p}$) \# Eq.~\ref{eq:m} \label{alg:mcmc-ga:line:accept} \\
                    \If{accepted} {
                        $confs_{accepted}.add(conf_p)$ \label{alg:mcmc-ga:line:add_accepted} \\
                    \If{$perf_p > perf_{best}$} { \label{alg:mcmc-ga:line:update_best_start}
                        $conf_{best}, perf_{best} \leftarrow conf_p , perf_p$ \\  
                    } \label{alg:mcmc-ga:line:update_best_end}
                    } 
            }   \label{alg:mcmc-ga:line:loop_end}
            $\Delta perf$ $\leftarrow$ $(perf_{best} - perf_{seed})/perf_{seed}$ \label{alg:mcmc-ga:update_global_imp}\\
            $conf_{parents} \leftarrow evolve(conf_{best}, confs_{accepted})$  \label{alg:mcmc-ga:line:evolve} \\
            $generation \leftarrow generation+1$ \label{alg:mcmc-ga:update_gens}\\
         } \label{alg:mcmc-ga:line:convg}
         \Return{$conf_{best}$}
\end{algorithm*}
\endgroup}
\begin{document}

\title{\textsc{ConEx}: Efficient Exploration of Big-Data System Configurations for Better Performance}

\author{Rahul Krishna$^{\dagger}$,~Chong Tang$^{\dagger}$,~Kevin Sullivan,~and~Baishakhi Ray
\IEEEcompsocitemizethanks{
\IEEEcompsocthanksitem $^{\dagger}$Krishna, R., and Tang, C., are joint first authors.
\IEEEcompsocthanksitem Tang, C. is with Walmart Labs, Mountain View, CA USA.\protect\\
E-mail: ct4ew@virginia.edu
\IEEEcompsocthanksitem Sullivan, K. is with the Department of Computer Science, University of Virginia, Charlottesville, VA USA.\protect\\
E-mail: sullivan@virginia.edu
\IEEEcompsocthanksitem Ray, B. and Krishna, R. are with the Department of Computer Science, Columbia University, New York City, NY USA.\protect\\
E-mail: rayb@cs.columbia.edu and i.m.ralk@gmail.com}
}
\markboth{IEEE Transactions on Software Engineering, submitted September `19}{Chong \etal: ConEx: Efficient Exploration of Big-Data System Configurations for Better Performance}


\IEEEtitleabstractindextext{%
\begin{abstract}
    Configuration space complexity makes the big-data software systems hard to configure well. Consider Hadoop, with over nine hundred parameters, developers often just use the \textit{default} configurations provided with Hadoop distributions. The opportunity costs in lost performance are significant. Popular learning-based approaches to auto-tune software does not scale well for big-data systems because of the high cost of collecting training data. We present a new method based on a combination of \textit{Evolutionary Markov Chain Monte Carlo (EMCMC)} sampling and cost reduction techniques to find better-performing configurations for big data systems. For cost reduction, we developed and experimentally tested and validated two approaches: using scaled-up big data jobs as proxies for the objective function for larger jobs and using a dynamic job similarity measure to infer that results obtained for one kind of big data problem will work well for similar problems. Our experimental results suggest that our approach promises to improve the performance of big data systems significantly and that it outperforms competing approaches based on random sampling, basic genetic algorithms (GA), and predictive model learning. Our experimental results support the conclusion that our approach strongly demonstrates the potential to improve the performance of big data systems significantly and frugally.
\end{abstract}

}

\maketitle
\section{Introduction}
\label{sect:introduction}

The use of Big-data frameworks such as Hadoop and Spark has become a de-facto standard for developing large scale data-driven applications. These frameworks are highly configurable and can be tailored to meet a diverse set of needs. 
In practice, however, it is hard to fully exploit such configurability. Instead, off-the-shelf, or \textit{default}, configurations are most commonly used~\cite{ren2013hadoop}.  This often leaves significant performance potential unrealized~\cite{xu2015hey}. Configuring for performance is important especially for big data because~``even a small performance improvement translates into significant cost savings due to the scale of computations involved~\cite{jia2016auto}''.

Finding high-performing (or \textit{good}) configurations for big data systems is hard. Their configuration spaces are vast and their configuration-to-performance functions are complex. 
First of all, they have multiple configurable sub-systems~\cite{white2012hadoop}. Hadoop, for example, has about 900 parameters across 4 sub-systems. Some parameters, e.g., numeric ones, have many values. 
Secondly, parameters also have diverse types, including optional and dependent substructures.  For example, setting one Boolean parameter to {\em true} can enable a feature, requiring values for all of its parameters. 
Further, parameters can also be constrained by external values. 
For example, in a multi-core system, one cannot set the  \textit{number-of-core} value to a number larger than the number of available cores. Also, due to its discrete nature, typical mathematical optimization methods do not apply~\cite{louviere2011design}. Finding good configurations ends up as a black-art~\cite{Joshi:2012:AHP:2188286.2188323}.

\begin{figure}[tp!]
    \setlength\abovecaptionskip{-0.18\baselineskip}
    \centering
    \includegraphics[width=\linewidth]{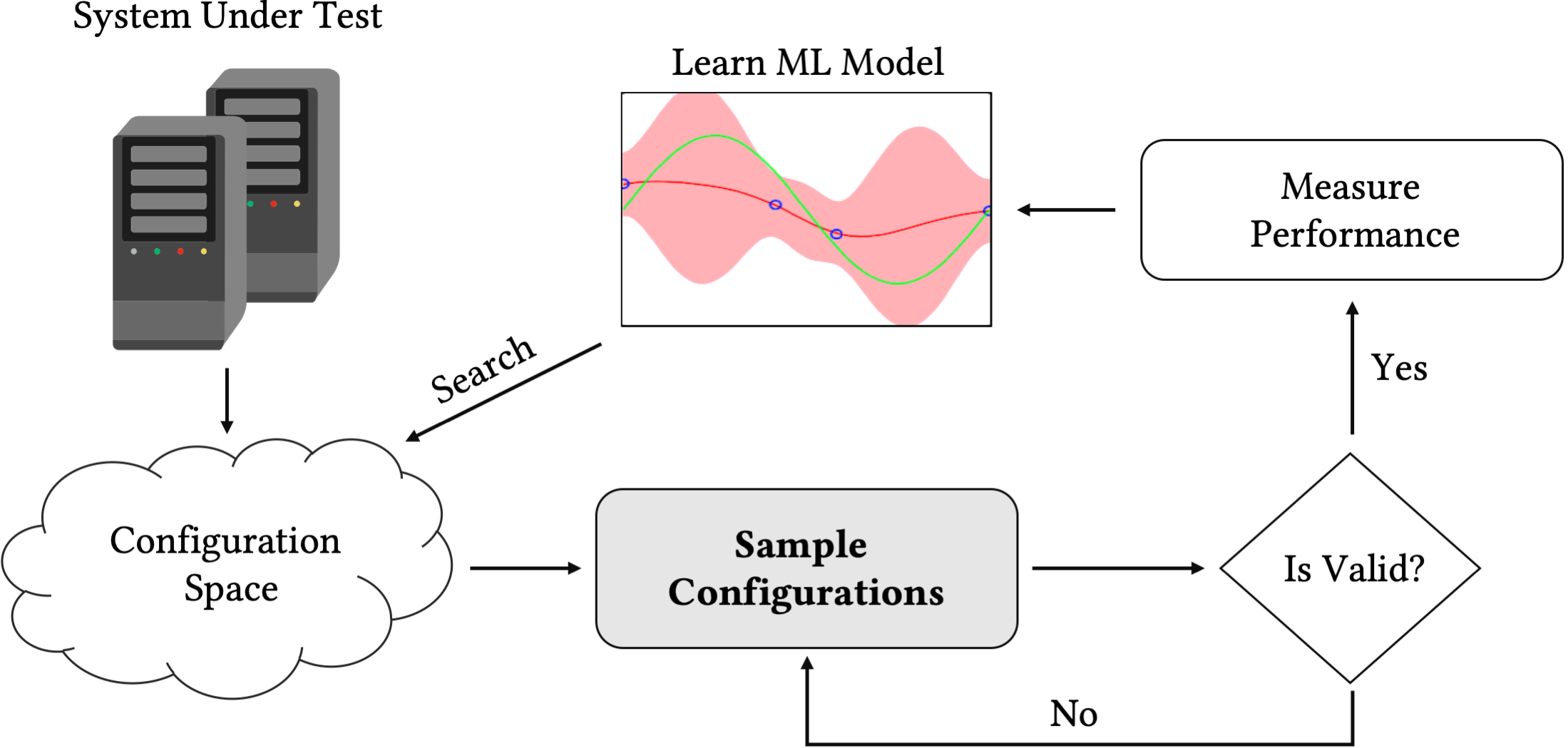}
    \caption{A typical framework for machine learning based automatic configuration of software sytems.}
    \label{fig:intro_fig}
\end{figure}

For traditional software, the problem of finding better configuration has given rise to a significant body of work~\cite{siegmund2012predictperf, siegmund2015performance, zhang2015performance, venkataraman2016ernest,sarkar2015cost,guo2013variability}. This research share a common general framework shown in \fig{intro_fig}. They involve the following steps: 
(i) deploy different sampling strategies to select a set of valid and representative configurations~\cite{gogate2006new,chakraborty2014distribution,lei2008ipog,johansen2012algorithm,marijan2013practical,kaltenecker2019distance},
(ii) use the sampled configurations to measure the performance of the system,
(iii) use Machine Learning (ML) models to learn the mapping between configuration and performance~\cite{siegmund2012predictperf,siegmund2015performance,guo2013variability,nair-bad-learners,nair2018finding}, and finally, 
(iv) use the learned models to find better configurations.
Most notably, \textbf{the success of these learning-based approaches is contingent on the size and the quality of the sampled configurations used for training}.

For big-data systems, existing techniques often struggle to scale due to two reasons. First, the cost of collecting training data in big data systems is prohibitively large~\cite{weiss2008maximizing}. For example, in a typical Hadoop system, a single run of a simple sorting task for a typical workload (data size=3.2GB) can take up to 14 minutes. Typical learning-based approaches need about 2000 samples~\cite{ye2003recursive}. For the sorting task in Hadoop, this would take 19 days to obtain! Therefore, we seek multiple orders of magnitude cost reduction for such a method to be practical. Second, The configuration space of big data systems is complex. The configuration-finding problem for big data systems involves a significantly larger number of dimensions than addressed in most previous work. Nair~\etal~\cite{nair-bad-learners} showed that for complex, configurable systems the measurement data often do not generalize well making it almost impossible to find the ``best'' configuration.

To address this issue, Nair \etal~\cite{nair-bad-learners} proposed a rank-based learning approach. Instead of focusing on constructing an accurate model, they use a random sampling strategy to build a ``bad'' model that can learn to predict whether one given configuration is better than another. They showed such predictors can be trained easily with significantly fewer samples. For big-data systems, as we will show in ~\tion{results}, it is hard to learn even such rank-preserving models with high accuracy. One might try Neural Networks (NNs)~\cite{ha2019DeepPerf}. However, NNs require large amounts of high-quality samples data for training, and the cost of even collecting such data for big-data systems would be prohibitively high~\cite{pavlo2009comparison}.

Given the number of samples needed to train a good model and the cost involved in collecting them for big-data systems, we cannot rely on the popular random sampling$+$learning based approaches. They must be eschewed in favor of better methods that (a)~can give us {\em near-optimal} configurations within a {\em sampling budget} and more importantly (b)~can be scaled much more easily.

To achieve these objectives, in this paper \textbf{we argue that random sampling is inadequate and we need smarter sampling strategies that can explore {diverse} configurations with a {high discriminatory power}}. In particular, we show that Evolutionary Markov Chain Monte Carlo (EMCMC) sampling strategy is best suited for this purpose. A nice property of EMCMC is that, unlike random sampling, it tries to estimate the target distribution, \ie draws more samples from important regions while maintaining sample diversity.

Overall, our work makes the following contributions:
\bi
\item We implement a configuration exploration framework for big-data systems called \tool. We demonstrate experimentally that \tool outperforms learning-based methods in the case of big-data systems.

\item We make available a replication package for \tool to accompany this paper at \href{https://git.io/Jv958}{git.io/Jv958}. 

\item We show that Evolutionary Markov Chain Monte Carlo (EMCMC) based strategy can be effective in finding the high-performing configurations in a complex and high-dimensional configuration space of big-data systems.  In general, EMCMC outperforms random and evolutionary (\ie genetic algorithm based) sampling strategies. 
\item We find compelling evidence that \tool can \textit{scale-up}, \ie, good configurations found with small workloads work well for significantly larger workloads and saves significant experimental cost. 

\item We find compelling evidence that \tool can \textit{scale-out}, \ie,  good configurations for one kind of job would yield improvements for other \textit{dynamically similar} jobs, saving further sampling cost.
\ei
The rest of this paper is organized as follows: 
\S~\ref{sect:background} provides background. \tion{mcmc} discusses MCMC. \S~\ref{sect:approach} introduces \tool. \S~\ref{sect:experiment} presents the experimental design and evaluation. \S~\ref{sect:results} highlights our experimental results. \S~\ref{sect:related} samples some previous related work in this area. \S~\ref{sect:threats} highlights some threats to the validity. \S~\ref{sect:conclusion} offers some concluding remarks.

\section{Formalization and Background}
\label{sect:background}

\subsection{Terminologies and Example}
\label{sect:terminologies}
The following are some of the frequently used terms:
\bi[wide=0pt]
\item Configuration {\it parameter}~$p_i$: Is a configuration variable whose value is set by a system installer or user to invoke certain desired system property.

\item Configuration \textit{type}~$t$: Is an $N$-element record type $[p_1, \ldots, p_N]$, where each element $p_i$ is a configuration parameter and $N$ is the number of parameters representing the dimensionality of the space.  
\item {\it Configuration} $c$: Is a configuration type, $t$, in which valid values are assigned to the configuration parameters $p_i$. 
\item Configuration {\em space} $\zeta$:  Is the set of all valid configurations for a given system. The definition of $valid$ varies from system to system. If there are no constraints on a configuration $c$, and if $N$ is the number of parameters and $M$ is the average number of possible values of each parameter, then the size of configuration space $\zeta$ is roughly equal to $M^N$. 
\item {\it Performance} $Y$: The measured performance of the software system given that it is configured according to $c$. A number of performance measurements can be made for a system. For example, we may want to \textit{maximize} performance measures such as \textit{throughput} or \textit{minimize} measures such as \textit{latency}. 
\item {Target distribution {\it P(Y~|~c)}: The conditional distribution that models the performance of the software system given the configuration c.}   
\ei

Table~\ref{tab:conf-tuple} presents two excerpts of sample configurations, $\mathbf{c_1}$ and $\mathbf{c_2}$, for Hadoop. In practice, configurations have hundreds of parameters, of varying types: Boolean, integer, categorical, string, etc. In total, as per \tab{studied-parameters}, Hadoop has 901 and Spark has 212 configuration parameters giving rise to $~3*10^{28}$ and $~4*10^{16}$ total configurations respectively.

\begin{table}[!hp]
\caption{Two Hadoop configuration examples}
\label{tab:conf-tuple}
\arrayrulecolor{black}
\centering
\vspace{1em}
\resizebox{0.825\linewidth}{!}{
    \begin{tabular}{@{}l|r|r@{}}
        \textbf{Parameters} & $\mathbf{c_1}$ & $\mathbf{c_2}$ \\ \hline
        dfs.blocksize & 3     &  2 \\
        mapred.job.ubertask.enable & FALSE & True \\
        mapred.map.java.opts   & -Xmx1024m & -Xmx2048m \\ 
        $\cdots$ &$\cdots$&$\cdots$\\
        mapred.shuffle.merge.percent & 0.66  & 0.75  \\
    \end{tabular}}%
\end{table}

\subsection{Identifying optimal configurations with Stochastic Approximation}
\label{sect:formal_current}
In the most general sense, the goal of finding the best configuration for a configurable software system can be understood as a search problem. The goal of this search problem would be to identify a configuration that maximizes (or minimizes) an objective function. More formally, given a space of all possible configurations $\zeta$, and valid configuration from that space $c$, let us represent the software system as a function $S: c \rightarrow Y$. The function $S$ consumes an input $c$ and returns the performance Y. The goal of finding the optimal configuration can be viewed as a search for a configuration $c^* \in \zeta$ such that we obtain the best performance $Y^*$. This can be generalized as follows:
\begin{equation}
    \left\{ {\begin{array}{*{40}{l}}
        \\[-0.95em]
        {Y^* = \mathop {\min }\limits_{c \in \zeta } Y \equiv \mathop {\min }\limits_{c \in \zeta } S(c)}&{Y = {\rm{Latency, etc.}}}\\[0.5em]
        {Y^* = \mathop {\max }\limits_{c \in \zeta } Y \equiv \mathop {\max }\limits_{c \in \zeta } S(c)}&{Y = {\rm{ Throughput, etc.}}}
        \end{array}} \right.
\end{equation} 
However, the space of configurations, $\mathcal{\zeta}$, is exponentially large. Further, the software system $S$ is quite complex and each of the configuration $c\in\mathcal{\zeta}$ is highly nonlinear, high dimensional, and are otherwise inappropriate for deploying deterministic optimization strategies. Therefore, we seek alternative strategies to find optimum configurations~\cite{oh2017near-optimal}.

A prominent alternative to deterministic optimization is the use of machine learning models in conjunction with stochastic optimization methods to solve the search problem of finding the optimum configuration for a given software system~\cite{siegmund2015performance,siegmund2012predictperf,nair-bad-learners}. These methods use a three-step approach described below: 

\noindent(i)~\textit{Stochastic Sampling}. This step attempts to overcome the issue of exponentially large space of possible configurations $c\in\zeta$. The sample of configurations $c$ are drawn from an underlying probability distribution $f(c) = f(p_1, p_2, ..., p_N)~\forall~c\in\zeta$. This distribution is almost always assumed to be uniform in nature, \ie, $f_\zeta(c) \rightarrow \mathcal{U}(p_1, p_2, ..., p_N)~\forall~c\in\zeta$. The total number of samples that are drawn are limited by a sampling budget.

\noindent(ii)~\textit{Modeling with machine learning}. Even with a limited number of samples, the time and cost overhead of having to measure the true performance can be exorbitant. Therefore, from among the sampled configurations, a few representative samples are used to construct a machine learning model to approximate the behavior of the software system. More formally, the machine learning model can be represented as a function $ML$ that takes as input a configuration $c$ and returns an estimated performance measure $\hat{Y}$, \ie, $ML: c \rightarrow \hat{Y}$ where $ML$ is a machine learning model and $\hat{Y}$ is the performance predicted by the ML model. 
 
\noindent(iii)~\textit{Identifying the best configuration}. With the machine learning model from above, the performances of the remaining configuration samples are predicted. The configuration yielding the best performance is returned as being the optimum setting for the given software system.  

\subsubsection{Challenges with the current state-of-the-art}
There are several challenges associated with using the methodology discussed above. Foremost among them is due to the stochastic sampling step. Since this sampling step precedes the construction of an ML model, the quality of samples directly affects the subsequent steps. 

The problem with sampling arises due to the assumption that the configuration parameters $p_i$ follow a uniform distribution. This assumption is fraught with risks\textemdash 
\be
\item Most of the time modeling might be spent exploring sub-optimal regions of the configuration space $\zeta$.
\item The machine learning model build with these samples tend to be severely biased. As a result, they often fail to identify the best configuration if it exists in the regions of the configuration space that are previously unseen.
\item The optimum configuration identified by the ML model is at best only an estimate of the local optima.   
\ee

    The aforementioned problems can be a major limiting factor for big data systems such as Hadoop and Spark where both the configuration space and the cost of dynamic sampling are often significantly large. Also, it usually takes a longer time to run big data jobs; hence, generating enough dynamic samples within a limited exploration budget is challenging. 
    Thus, getting stuck to some local region is a common problem. This paper aims to address these issues with Evolutionary Markov Chain Monte Carlo (EMCMC) based stochastic search strategy, as discussed next.

\section{Evolutionary Markov Chain Monte Carlo}
\label{sect:mcmc}

\noindent
The problem of finding optimal configuration would be trivial if one knew how the configurations ($c$) affect the performance of the software system ($Y$), \ie, if the distribution $P(Y~|~c)$ was known beforehand. When $P(Y~|~c)$ is unknown, one workaround could be to randomly draw enough samples from the configuration space to approximate the function. However, given the vastness of the configuration space of the big data system, a random draw is unlikely to find an optimal solution without some additional guidance. To overcome this, we use a class of algorithms based on Evolutionary Markov Chain Monte Carlo (EMCMC). While MCMC based methods are used to generate configurations from the unknown distribution $P(Y~|~c)$, the evolutionary component provides some additional guidance towards reaching the optima faster.

\begin{figure}[bt!]
    \setlength\abovecaptionskip{-0.05\baselineskip}
    \setlength\abovecaptionskip{-0.05\baselineskip}
    \centering
    \includegraphics[width=0.9\linewidth]{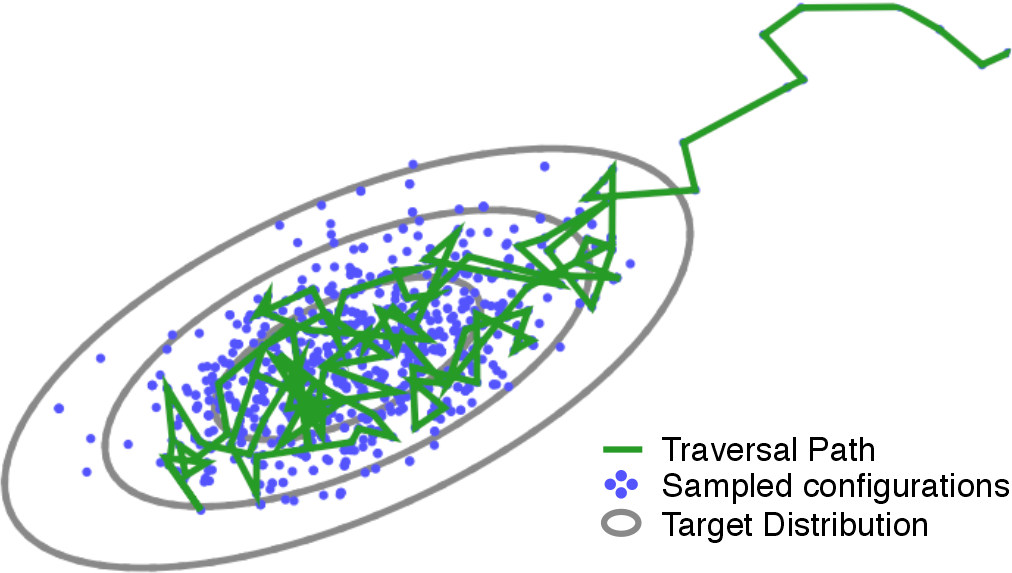}\\[0.5em]
    \caption{An example of \textit{MCMC} traversal. The {\color{fig1_green}\textbf{green line}} represents the state transition paths of the MCMC algorithm, the {\color{fig1_blue}\textbf{blue dots}} represents the samples and the {\color{gray80} \textbf{gray ellipses}} represent the boundaries of the target distribution.}
    \label{fig:mcmctraversal}
\end{figure}

MCMC works by constructing a Markov Chain to generate a sequence of configuration samples where a new sample ($c_{t+1} \rightarrow Y_{t+1}$) depends only on its previous sample ($c_{t} \rightarrow Y_{t}$).  The process of generating a next sample given a current sample is called a state-transition. We approximate the unknown distribution by recording all the state transitions; as the number of state transitions increases the distribution of the new samples converge more closely to the desired distribution (as illustrated in \fig{mcmctraversal}).

The rest of this section discusses the use of MCMC to find optimal configurations. Specifically, our preferred MCMC algorithm (Metropolis-Hastings Algorithm) is discussed in~\tion{mh}, the evolutionary variant used for optimization (EMCMC) and its benefits are explained in~\tion{emcmc}.

\subsection{Metropolis-Hastings Algorithm}

\label{sect:mh}

\noindent
A number of MCMC based methods exists in literature~\cite{ritter1992facilitating, hastings1970monte, goodman2010ensemble, tibbits2014automated}. From these, we choose the Metropolis-Hastings algorithm~\cite{metropolis1953equation, hastings1970monte} as they are best suited for deriving samples from high-dimensional spaces such as those observed in configuring big-data systems. 

Given a configuration $c=\left[p_1, p_2, p_3, ..., p_N\right]$, with $N$ configurable parameters denoted by $p_i$ and the performance measure $Y$ (\eg execution time), there exists an unknown conditional distribution function $P(Y|c)$ that informs us of the performance $Y$ of the job under test given a configuration $c$. However, without exploring the entire configuration space, $P(Y|c)$ cannot be inferred.   

The Metropolis-Hastings algorithm attempts to generate a new configuration from the unknown distribution function $P(Y|c)$ using an \textit{approximate function}  $Q(Y|c)$ to draw samples, where $Q(Y|c)$ is proportional. but may not be identical, to the original distribution. 

From $Q(Y|c)$, the Metropolis-Hastings algorithm generates a sequence of configurations such that the probability of selecting the next configuration  ($c_{t+1}$) is dependent only on the current configuration ($c_{t}$).  
Since the configuration $c_{t+1}$, which is generated at step $t+1$, depends only on its immediate predecessor $c_{t}$, the sequence of samples belong to a first order Markov chain.

At each step, the algorithm measures an \textit{Acceptance Probability} $A(c_{t+1}|c_{t})$ that determines if the newly sampled candidate ($c_{t+1}$) will be accepted or rejected. The acceptance probability compares the performance of the newly  generated candidate (\ie, $Q(Y|c_{t+1})$) with respect to where the current sample lies (\ie, $Q(Y|c_{t})$). It is given by:
\begin{equation}
    \label{eq:m}
    A(c_{t+1}|c_{t}) = min\left(1,\frac{Q(Y|c_{t+1})}{Q(Y|c_{t})}\right)
\end{equation}

Based on the acceptance probability the new candidate configuration is either accepted (in which case the candidate value is used in the next iteration) or rejected (in which case the candidate value is discarded, and the current value is reused in the next iteration).

The key component of the acceptance probability is that it is determined by the ratio $Q(Y|c_{t+1})/Q(Y|c_{t})$. If the new configuration $c_{t+1}$ is {\it more} likely to produce a better performance than the current configuration $c_t$, then $Q(Y|c_{t+1})\!>\!Q(Y|c_t)$ and 
$Q(Y|c_{t+1})/Q(Y|c_{t})\!>\!1$. Consequently, according to \eq{m}, $A(c_{t+1}|c_{t})=1$ and we will always accept the the new configuration. 

On the other hand, if the new configuration $c_{t+1}$ is {\em less likely} to produce a better configuration than the current configuration, then $Q(Y|c_{t+1})\!<\!Q(Y|c_t)$ and $Q(Y|c_{t+1})/Q(Y|c_{t})\!<\!1$. In this case, $A(c_{t+1}|c_{t})$ is less than 1 and we will sometimes reject the new configuration. However, since the acceptance probability $A(c_{t+1}|c_{t})$ is non-zero, depending on the probability value, we may sometimes accept new configurations that perform worse than the previous configuration. The poorer the new configuration performs, the smaller the value of $A(c_{t+1}|c_{t})$ will be, thereby making it less likely for us to accept a configuration with a very bad performance score. Using acceptance probability maintains the diversity among newly generated samples instead of always greedily choosing the better performants, and thus, slowly moves towards the target distribution. 





The MCMC algorithm is designed to spend most of its time in the high-density region of the target distribution. Consequently, the samples of configurations obtained using MCMC are highly likely to contain the global optima among them.

\subsection{EMCMC: An Evolutionary Extension to MCMC}
\label{sect:emcmc}

Traditionally, MCMC uses some arbitrary distribution to generate the next sample configuration $c_{t+1}$, given the current sample $c_{t}$. The initial candidate configurations are mostly non-performant and are often discarded. Thus, although MCMC converges to the global optima eventually, it does so very slowly. This is detrimental because we may quickly exhaust our computational budget before finding a high-performant configuration.

To expedite the convergence, we ought to modify the existing MCMC to offer some additional guidance during the generation step in the Metropolis-Hastings algorithm discussed above (\ie, Step-1). In this paper, we propose a novel \textit{Evolutionary-MCMC} algorithm (or \textit{EMCMC} for short). Like evolutionary algorithms (such as genetic algorithms~\cite{holland1992adaptation,akbari2011multilevel,whitley1994genetic}), in EMCMC we start with an initial {\em population} of $N > 1$ configurations. We then repeat the following steps until the allocated budget is exceeded: 
\be 

\item \textbf{Evolutionary Generation}: {We randomly choose a subset of initial configurations}. For each configuration in the subset ($c_t$), we \textit{generate} a new configuration $c_{t+1}$ by applying (a)~mutation; and (b) cross-over operations. That is:
    \bi[wide=0pt]
        \item \textit{Mutation}: As described in \tion{terminologies}, a configuration $c$ is comprised many parameters $p_i$. During mutation, some of these parameters are randomly changed (with allowed values) to form a new configuration, say $c_{t+1}$.
        \item \textit{Cross-over}: We randomly selecting two parent configurations, from among all the $c_t^i$ and $c_t^j$. Each configuration is bisected, \ie, divided in 2-parts. Then the first part of $c_t^i$ is spliced with the second part of $c_t^j$ and vice versa, generating two offspring configurations $c_{t+1}^i$ and $c_{t+1}^j$.
    \ei

\item \textbf{Acceptance or Rejection}: For each new configuration generated with the previous step, we compute the acceptance probability according to Equation~\ref{eq:m}. 

\item \textbf{Transition}: Using the acceptance probability for each $c_{t+1}$, we either retain those configurations, or we reject them. All the retained configurations represent the next state of the EMCMC algorithm. 
\ee

\subsubsection{Generalizability of EMCMC}

\noindent

    The rest paper uses EMCMC strategy as part of the \tool framework to generate new samples. It is worth noting that EMCMC can be applied to any domain where one not only needs to generate samples that belong to an unknown distribution but also requires the newly generated samples to exhibit some desired property (such as optimizing performance). Accordingly, EMCMC based approaches offer some marked benefits over both traditional MCMC and genetic algorithms. 

\textit{Benefits over traditional genetic algorithms.} 
Ordinary genetic algorithms are susceptible to getting trapped at local optima~\cite{srinivas1994genetic}. This is because, in regular genetic algorithms, a new configuration will never be accepted even if its performance is only a little worse than that of its parents and any worse configuration will always be discarded. In contrast, the acceptance or rejection of new configurations in EMCMC is contingent on the acceptance probability which sometimes accepts inferior configurations. This avoids a quick convergence to local optima and increases the chances of eventually reaching the global optima. 

\textit{Benefits over traditional MCMC.} As mentioned previously, pure MCMC based optimization algorithms converge very slowly. In contrast, by using principles of evolutionary algorithms, EMCMC ensures that MCMC has some guidance as it approaches the global optima.

\section{\tool: Configuration Explorer}
\label{sect:approach}



 
Our approach is to use an EMCMC algorithm to sample Hadoop and Spark configuration spaces in search of high-performing configurations. We have implemented our approach in a tool called \tool. The rest of this section describes our approach in detail.

\begin{figure}[t]
    \centering
    \includegraphics[width=\linewidth]{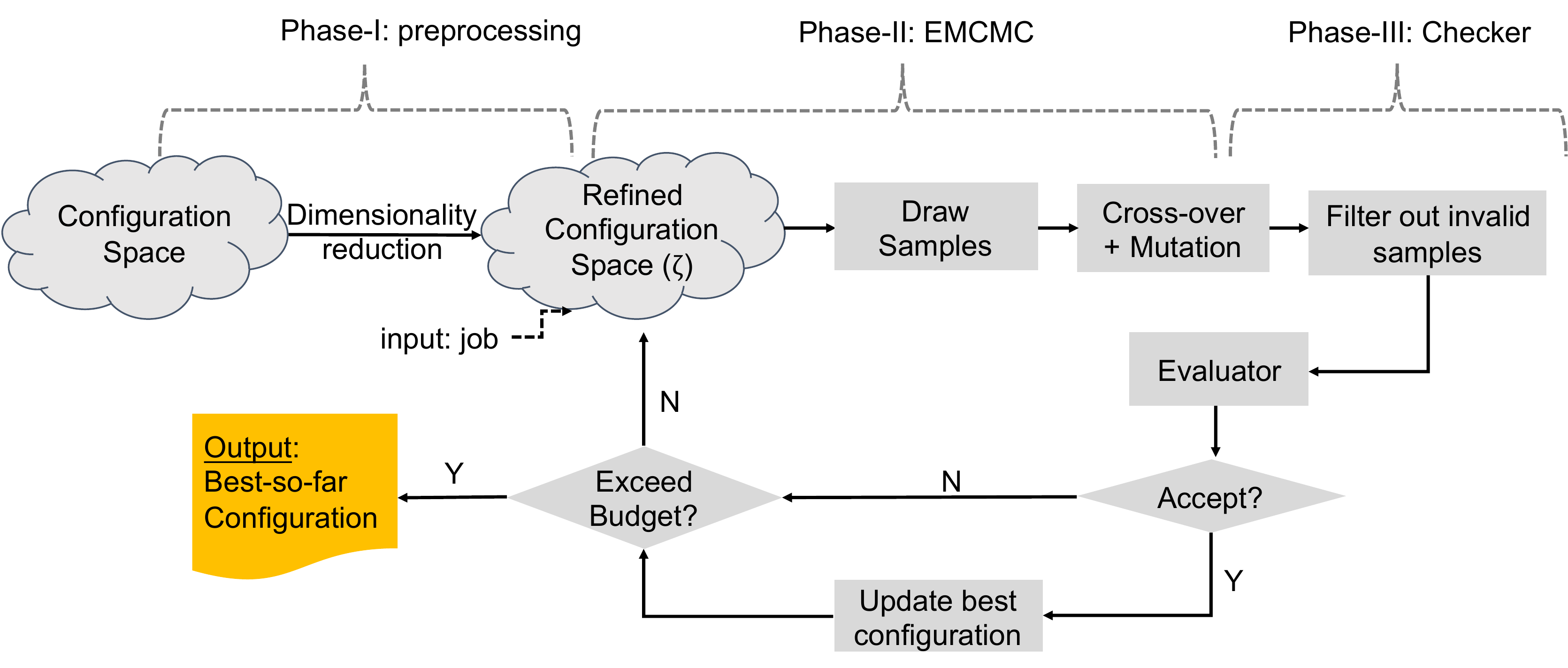}
    \caption{Workflow of the \tool framework.}
    \label{fig:framework}
\end{figure}

Figure~\ref{fig:framework} presents an overview of \tool. 
\tool takes a big data job as input and outputs the best configuration it found before exceeding a sampling cost budget. 
\tool works in three phases. 
In Phase-\rom{1}, it reduces the feature space by filtering out the configuration parameters that are not relevant to performance. 
In Phase-\rom{2}, it uses EMCMC sampling to find better configurations. Sampling starts with the default system configuration as the seed value. While sampling, \tool discards invalid configurations generated during sampling using a checker developed by Tang et al. \etal~\cite{tang2017interpreted} (Phase-\rom{3}).  If a configuration is valid, \tool runs a benchmark job using it and records the CPU time of the execution. It then compares the result with that of the best configuration seen so far, updating the latter if necessary, per our acceptance criterion (see~\Cref{eq:m}). 
Accepted configurations are subjected to cross-over and mutation, as described in~\Cref{sect:emcmc}, to produce configurations for the next round of sampling.  Once \tool exceeds a pre-set sampling budget, it outputs the best configuration found so far. We now describe each of these steps in greater detail.

\subsection{Phase-I: Pre-processing the configuration space}  
Hadoop and Spark have $901$ and $212$ configuration parameters, respectively (Table~\ref{tab:studied-parameters}). Yet most do not affect performance. In this step, we reduce the dimensionality of the configuration space by filtering out the parameters that do not affect the performance\textemdash this is similar to standard feature selection technique in Statistics or Machine Learning~\cite{guyon2003introduction}. 
We reduce the dimension two ways: we consider only parameters relevant to performance using our domain knowledge; and we select only a few values for sampling for each  parameter.

\begin{table}[b!]
\setlength{\tabcolsep}{5pt}
\setlength\abovecaptionskip{0.8\baselineskip}
\setlength\belowcaptionskip{-0.2\baselineskip}
\caption{{Configuration Space Characteristics}}
\label{tab:studied-parameters}
\centering
\resizebox{\linewidth}{!}{
\begin{tabular}{l|c|cccccc|c}
\toprule
\textbf{System} & \rotatebox{90}{\textbf{Total}}~\rotatebox{90}{\textbf{Parameters}~}~& \rotatebox{90}{\textbf{Total}} & \rotatebox{90}{\textbf{Bool}} & \rotatebox{90}{\textbf{Int}} & \rotatebox{90}{\textbf{Float}} & \rotatebox{90}{\textbf{Categorical}} & \rotatebox{90}{\textbf{String}} & \rotatebox{90}{\textbf{Approx.}}\rotatebox{90}{\textbf{Total}}\bigstrut\\\hline
\textbf{Hadoop v2.7.4} & 901   & 44     & 4     & 26    & 6     & 3 &   5 & $~3*10^{28}$ \bigstrut\\
\textbf{Spark v2.2.0}  & 212   & 27     & 7     & 14    & 4     & 2 &   0 & $~4*10^{16}$ \bigstrut\\
\bottomrule
\end{tabular}
}
\end{table}
The first part is manual, and based on a study of technical manuals and other work~\cite{Bonifacio:research_review}. For example, we removed Hadoop parameters related to version (\eg, java\-.runtime\-.version),  file paths (\eg, hadoop\-.bin\-.path),  authentication (\eg, hadoop\-.http\-.authentication\-.kerberos\-.keytab), server-address, and ID/names (\eg, mapreduce.output.basename). For Spark, we selected parameters related to the runtime environment, shuffle behavior, compression and serialization, memory management, execution behavior, and scheduling. 
In general, we err on the side of caution-- to make sure that we cover all parameters related to runtime performance, we select all parameters that have somewhat impact on the final performance. If we are not sure about a parameter, we include it in the sampling space.
{\color{black} Overall, the selecting the features during the pre-processing required a few man-hours worth of manual inspection. After feature subset selection, we ended up with a total of 44 and 27 parameters for Hadoop and Spark respectively (summarized in table~\ref{tab:studied-parameters}) that may possibly impact the performance. The remaining parameters are related to versions, paths, etc. and they tend not to affect the performance.}

We then finitely sub-sampled the ranges of integer, float, and string types. In particular, we sub-sample the configuration space by defining small ranges of values for each parameter, varying by parameter type. Boolean parameters are sampled for true and false values. Numerical parameters are sampled within a certain distance from their default. 
Even these reduced configuration spaces are several orders of magnitude larger than those studied in previous work. For example, most systems studied by Nair et al.~\cite{nair-bad-learners} have only thousands or at most a few million configurations. 
Table~\ref{tab:studied-parameters} summarizes the resulting configuration spaces that we dealt with for Hadoop and Spark.

\subsection{Phase-II: Finding better configurations}

This phase is driven by an EMCMC sampling strategy and implemented by~\Cref{alg:mcmc-ga,alg:evolution}.
\Cref{alg:mcmc-ga} is the main driver;~\Cref{alg:mcmc-ga:line:input} lists inputs and outputs. The algorithm takes a reduced configuration space ($\zeta$) and a given job as inputs. \tool samples configurations from $\zeta$ and evaluates their performance \wrt the input job. 
The routine also requires a seed configuration ($conf_{seed}$), and a termination criterion based on a maximum number of generations ($max\_gen$). We choose $max\_gen=30$ in our experiment. The tool then outputs are the best configuration found ($conf_\text{best}$) and its performance ($\mathit{perf}_\text{best}$).

\Cref{alg:mcmc-ga:line:init_perf,alg:mcmc-ga:init_gens,alg:mcmc-ga:init_global_imp} initialize some parameters including setting the best configuration and performance to the respective seed values.  ~\Cref{alg:mcmc-ga:line:sample_fst_generation} gets the first generation of configurations by randomly sampling $n$ items from  $\zeta$. 
We choose $n=4D$ where D is the number of parameters, but it could be set to any reasonable value. Lines~\ref{alg:mcmc-ga:line:loop_start} to \ref{alg:mcmc-ga:line:loop_end} are the main procedure for evaluating and evolving \wrt each configuration. 
Given a configuration $conf_p$,~\Cref{alg:mcmc-ga:line:evaluate_confp} records the job's performance ($\mathit{perf}_p$) and~\Cref{alg:mcmc-ga:line:accept} decides whether to accept it based on~\Cref{eq:m}. 
If accepted, Line~\ref{alg:mcmc-ga:line:add_accepted} stores the accepted configuration to a list $confs_{accepted}$, which later will be used in generating next-generation configurations (\Cref{alg:mcmc-ga:line:evolve}).
If the accepted one is better than the best previously found,  Lines~\ref{alg:mcmc-ga:line:update_best_start} to~\ref{alg:mcmc-ga:line:update_best_end} update the state accordingly. 

Once all configurations in the first generation are processed, ~\Cref{alg:mcmc-ga:update_global_imp} computes the performance improvement achieved by this generation \wrt the seed performance. 
Next in~\Cref{alg:mcmc-ga:line:evolve}, the algorithm prepares to enter the next generation by generating offspring configurations using cross-over and mutation operations (see~\Cref{alg:evolution}). ~\Cref{alg:mcmc-ga:update_gens} updates the generation number. 
This process repeats until the termination criterion is satisfied (\Cref{alg:mcmc-ga:line:convg}). 
Finally, the last line returns the best found configuration and the corresponding performance.

Algorithm~\ref{alg:evolution} is the evolution sub-routine of the EMCMC algorithm, as described in~\Cref{sect:emcmc}. 
For preparing configurations of the next generation, it takes the best configuration found so far and a list of parent configurations as inputs. There are two main steps: cross-over and mutation. From the best configuration, ~\Cref{alg:evolve:get_crossover_parameters} selects half of all parameters as cross-over parameters ($P_{crossover}$), and~\Cref{alg:evolve:get_mutate_parameters} identifies 6\% of parameters as mutation parameters ($P_{mutate}$). Next, for each parent configuration,~\Cref{alg:evolve:crossover} exchanges the values of same parameters in two parents with $P_{crossover}$. Note that since the values of the same parameter is exchanged, their types are automatically preserved.
It then randomly mutates the values of the mutation parameters at~\Cref{alg:evolve:mutate}. The resulting offspring is added into the children set at~\Cref{alg:evolve:add_into_children}. A set of new offspring configurations is returned at~\Cref{alg:evolve:return}.  

\begin{figure}
\begin{minipage}{\linewidth}
    \algoemcmc
    \algoevolve
\end{minipage}
\end{figure}

\subsection{Phase-III: Configuration Validity Checking}
\label{sect:phthree}

Configuration spaces are often subject to constraints on one or more parameters. For example, Hadoop's \textit{JVM options} parameter is of \textit{string} type, but not any string value will work. 
These constraints must be met to generate valid configurations.  
This is typical in domains such as software product line optimization~\cite{sayyad2013scalable}, where we have access to the constraints in the form of feature models. Unfortunately, for big-data systems, such as Hadoop and Spark, such validity constraints are neither well documented nor strictly enforced. Thus, the likelihood to making configuration mistakes is increased. 

For big data system such mistake is costly as it increases the cost of dynamic sampling.  
In a previous work, we\footnote{Tang, C and Sullivan, K} have developed and employed a configuration constraint checker developed with COQ theorem prover~\cite{barras1997coq} to express and check constraints~\cite{tang2017interpreted}.
In this work we extend the open-source instrumentation\footnote{\href{https://github.com/ChongTang/SoS\_Coq}{https://github.com/ChongTang/SoS\_Coq}} of the checker to ensure that the newly generated constraints are valid. 

The checker provides expressive means for documenting properties and constraints on the configuration parameters, and the type checker checks that all constraints are satisfied. We express all the configuration constraints in the form of COQ rules. Note that, 
as we leverage an existing checker, here we just need to write the rules, which requires manual effort of a single day. 
Such expressiveness cannot be offered by just the documentation. For example, Hadoop informally documents (but does not enforce) that certain configuration parameter values should be multiple of the hardware page size. Our checker generates a type error if a violation occurs. 

\subsection{Using \tool in Production scale jobs}
The proposed framework for \tool may operate with any workload size. However, in order to extend its usability to production scale jobs, we employ the following two strategies:
\begin{enumerate}[leftmargin=*]
    \item 
    \textbf{Scale-up}: For cost-effectively sampling performance as a function of configuration during sampling processes, we run Hadoop and Spark jobs using inputs that are several orders of magnitude smaller (here, 100X) than those we expect to run in production. 
    \item 
    \textbf{Scale-out}: To amortize the cost of sampling and profiling, we use a dynamic similarity measure of big data jobs to decide when good configurations found for one kind of job might be used for another kind without any additional sampling activity. 
\end{enumerate}
The above strategies do not form part of the core \tool framework. Instead, they are used when \tool in deployed in a production environment.

\section{Experimental Design}
\label{sect:experiment}

We implemented \tool with about $4000$ lines of Python code\footnote{Replication package is available at \url{https://git.io/Jv958}}. Our experiments were conducted in our in-house Hadoop and Spark clusters. Each had one master node and four slaves, each with an Intel(R) Xeon(R) E5-2660 CPU, 32GB memory, and 2 TB local SSD storage. We assigned 20GB of memory for Hadoop on each node in our experiments. We also made sure that no other programs were running except core Linux OS processes. 

\subsection{Study Subject and Platform} 

Table~\ref{tab:studied-parameters} summarizes the parameters and their types that we have studied for Hadoop $v2.7.4$ and Spark $v2.2.0$.

To evaluate \tool, we selected big-data jobs from HiBench~\cite{huang2010hibench}, a popular benchmark suite for big data frameworks. It provides benchmark jobs for both Hadoop and Spark. For Hadoop, we selected five jobs: WordCount, Sort, TeraSort, PageRank, and NutchIndex. Of these, 
\texttt{nutchindex} and \texttt{pagerank} are used in websearch; \texttt{sort} and \texttt{terasort} sort data; and finally \texttt{wordcount} operates on text data. These jobs only need Hadoop to execute. 
For Spark, we selected five Spark jobs: WordCount, Sort, TeraSort, RF, and SVD. Here, \texttt{svd} and \texttt{rf} are machine learning jobs that are unique to Spark.

HiBench has six different sizes of input workload for each type of job, from ``tiny'' (30KB) to ``Bigdata'' (300GB). For our experiments, we used ``small (320MB)", ``large (3.2GB)'', and ``huge (32GB)'' data inputs. We ignore the \texttt{tiny} workloads since the size (32kb) is too small to accurately model I/O and memory requirements for larger workloads such as \texttt{large} and \texttt{huge}. We note that our smallest baseline workload (\texttt{small}) is 3MB and the largest workload (\texttt{huge}) is 3GB. Although, the memory and I/O requirements for \texttt{small} and \texttt{huge} are vastly different, they can still be accommodated within our hardware.

    \Cref{tab:evaluation-time} shows the CPU times taken by the Hadoop jobs running with default configurations. We used small inputs while sampling but then evaluated the resulting configurations using huge (100X larger) workloads.
    A HiBench execution has two steps: data preparation and job execution. 
We prepared the data manually to control the data used for each job. 

\subsection{Job Classification}
\label{sect:sim}

To test our {\em scale-out} hypothesis we needed a measure of job similarity. We settled on resource usage patterns rather than  HiBench job types for this purpose. HiBench classifies jobs by business purpose (\eg indexing and page rank jobs fall in \textit{Websearch} category), which does not necessarily reflect the similarity in resource usage patterns. Our approach is based on the profiling of run-time behavior using system call traces. Similar approaches have been widely used in the security community~\cite{khadke2012transparent, yoon2014toward, kasick2009system}. We use a Unix command-line tool, {\tt strace}, to capture system call traces for this purpose. 

Based on the system call traces, we represent each job by a four-tuple, $<A,B,C,D>$, where
$A$ is a system call sequence, 
$B$ is a set of unique string and categorical arguments across all system calls, 
$C$ expresses term frequencies of string and categorical arguments captured per system call, and  
$D$ is the mean value of the numerical arguments per system call.~\Cref{tab:call} shows an example tuple. 

\begin{table}[t]
\setlength\abovecaptionskip{0.5\baselineskip}
\setlength\belowcaptionskip{0.5\baselineskip}
\centering
\caption{Example Tuple representing resource usage of a job}
\label{tab:call}
\footnotesize
\setlength{\tabcolsep}{2pt}
\resizebox{\linewidth}{!}{
    \begin{tabular}{@{}cV{1}p{.62\columnwidth}@{}}
        \hlineB{2}
        Example System  &  $foo(1,\menquote{b}),bar(\menquote{b},True),$\bigstrut[t]\\
        Call Sequence      &  $foo(2,\menquote{b}),foo(3,\menquote{c})$\bigstrut[b]\\
        \hlineB{1}
        $A$ & $\{foo, bar, foo, foo\}$ \bigstrut\\
        $B$ & $\{foo: (\menquote{b}, \menquote{c}), bar:(\menquote{b})\}$ \bigstrut\\
        $C$ & $\{foo: [\menquote{b}: 0.66, \menquote{c}: 0.33], bar: [\menquote{b}: 1.0]\}$ \bigstrut\\
        $D$ & $\{foo: [\menquote{1^{st} arg}: 2.0]\}$ \bigstrut\\
        \hlineB{2}
    \end{tabular}}
\end{table}


To compute the similarity between two jobs, we calculate similarities between corresponding tuple-elements separately. Each contributes equally to the overall similarity measure. 
For $A$, we use pattern matching\textemdash 
we slice the call sequences and compute the similarity between them. 
To find the similarity between $B$ elements, we compute the Jaccard Index, which is a common approach to compute the similarity of two sets. 
For $C$ elements, we compute the average difference of each term frequency.  
Finally, for $D$ elements, we compute the similarity of mean value of numerical arguments as $1 - abs(mean1, mean2)/max(mean1, mean2)$. 
We take the average value of these four scores as the final similarity score between two jobs. 
We consider two jobs to be {\em similar} if their similarity score is above $0.77$ (\ie from third quartile (Q3) of all the similarity scores).

\subsection{Comparing with Baselines}
\label{sect:baseline}

We compare \tool's performance with three potentially competing approaches:
(i) a random sampling strategy, (ii) a genetic algorithm-based optimization strategy, and (iii) a learning-based approach. 
The first one evaluates whether EMCMC is a good sampling strategy, the second one checks EMCMC's ability to find a near-optimal configuration. The last one evaluates the choice of EMCMC over a model-learning approach. Here, we compare \tool with Nair \etal's ranking based approach~\cite{nair2017faster}, which is most relevant for our problem.

\subsection{Evaluation Criterion}
\label{sect:eval}

\noindent

Typically, performance optimization models are evaluated using \textit{performance gain}. 
Let the performances of a job with the default configuration be is $\mathit{perf}_\text{default}$, and the best discovered configuration be defined as $\mathit{perf}_\text{best}$. Then the performance gain $\Delta_\text{gain}$ measures the absolute percentage improvement in the performance. It is computed as below:
$$\Delta_\text{gain}\% = \left|\frac{\mathit{perf}_\text{default} - \mathit{perf}_\text{best}}{\mathit{perf}_\text{default}}\right| \times 100$$ 

Note that while it may make intuitive sense to compare $\mathit{perf}_\text{best}$ to the true-best configuration, the best configuration is unknown as the configuration space is extremely large. 
In contrast with machine learning based methods of Nair \etal~\cite{nair2017faster} that attempt to predict for the best configuration using a learning based approach, we use a evolutionary search 
to estimate the best configuration within a given budget. If we have infinite budget, \tool can theoretically converge on the true-best configuration. Therefore, we compare, the best configuration found within a given budget (\ie, $\mathit{perf}_\text{best}$) with the default configuration (\ie, $\mathit{perf}_\text{default}$).

\section{Experimental Results}
\label{sect:results}

\subsection*{RQ1.~\rqa}
The first research question seeks to provide a summary of the performance gains that can be achieved with the use of \tool for three workloads: small, large, and huge of Hadoop and Spark jobs. Specifically, we explore different configurations using \tool with smaller workloads (in RQ1-1); then we use the configurations obtained from the small workloads to check for performance gains of larger workloads, \ie, through scale-up hypothesis (in RQ1-2); and finally for different jobs that share similar dynamic characteristics (in RQ1-4). 

\noindent
\noindent\textbf{\textsf{\small RQ1-1.~How effective is \tool in finding optimal configurations for small workloads?}}
We investigate this RQ for Hadoop by exploring the configuration space using {\em small} workloads. HiBench generates a detailed report after each run, with performance information including CPU time. For Spark jobs, we intentionally refrained from using a small workload. This is because, for small workloads, the run-time in Spark was negligibly low compared to Hadoop. Therefore, to enforce a fair comparison, we used Spark ``large'' workloads, tailored to take as long as small jobs did in Hadoop (about 30 seconds per run). This ensures the cost of sampling is comparable between both systems. 

Our findings are tabulated in \tab{perf-improvement}. The best configurations found by \tool achieved between 7\% to 72\% improvements over the default configurations for five Hadoop jobs. On average, we notice a performance gain of 30.3\% for Hadoop jobs operating on a small workload. For Spark jobs, \tool produced 0.4\% to 40.4\% performance improvements for all five jobs with an average improvement of 10.6\%. Thus, we conclude that:

\begin{result}
    For Hadoop and Spark jobs with the small workload, \tool can find configurations that produce up to 72\% and 40\% performance gains over the default configurations in Hadoop and Spark respectively.
\end{result}

Even if \tool manages to find a better configuration with smaller workloads, \tool will be most effective if it can improve performance for larger workloads. It will obviate the need for making additional performance measurements thereby saving significant cost. 

\begin{table}
        \caption{{Performance improvement offered by \tool for \textit{Scale-up}. Note: the numbers reported below for scale-up were obtained using the Top-1 configuration from the baseline.}}
        \label{tab:perf-improvement}%
        \centering
        \textbf{\textsc{Hadoop}}\\[0.2em]
        \arrayrulecolor{black}
        \setlength{\tabcolsep}{4pt}
        \resizebox{\linewidth}{!}{
        \begin{tabular}{@{}rrrcc|r@{}}
        \hlineB{2}
        \multicolumn{6}{c}{\textbf{Small (Baseline)}} \bigstrut\\
        \hlineB{2}    
        \multicolumn{1}{c}{WordCount} & \multicolumn{1}{c}{Sort}  & \multicolumn{1}{c}{TeraSort} & \multicolumn{1}{c}{NutchIndex} & \multicolumn{1}{c}{PageRank} & \multicolumn{1}{c}{Average}\bigstrut\\\hline
        12.50\% & 72.10\% & 27.40\% & ~7.09\% & 32.70\% & 30.30\% \bigstrut\\
        \hline
        \multicolumn{6}{c}{\textbf{Small}~$\longrightarrow$~\textbf{Large}}\bigstrut\\\hline
        15.60\% & ~7.70\% & 16.00\% & 18.70\% & 44.70\% & 20.50\% \bigstrut\\\hline
        \multicolumn{6}{c}{\textbf{Small}~$\longrightarrow$~\textbf{Huge}}\bigstrut\\\hline
        21.50\% & 15.80\% & 18.30\% & 14.20\% & 25.20\% & 19.00\% \bigstrut\\\hlineB{2}
        \end{tabular}}\\[0.2em]
        \textbf{\textsc{Spark}}\\[0.2em]
        \resizebox{0.83\linewidth}{!}{
        \begin{tabular}{rrrcc|r}
        \hlineB{2}
        \multicolumn{6}{c}{\textbf{Large (Baseline)}} \bigstrut\\
        \hlineB{2}    
        \multicolumn{1}{c}{WordCount} & \multicolumn{1}{c}{Sort} & \multicolumn{1}{c}{TeraSort} & \multicolumn{1}{c}{RF} & \multicolumn{1}{c}{SVD} & {Average}\bigstrut\\\hline
        ~2.70\% & ~3.28\% & 40.41\% & ~6.35\% & ~0.38\% & 10.60\% \bigstrut\\\hline 
        \multicolumn{6}{c}{\textbf{Large} $\longrightarrow$ \textbf{Huge}} \bigstrut\\\hline
        ~5.80\% & ~1.60\% & 16.80\% & ~7.10\% & ~1.90\% & ~6.60\% \bigstrut\\
        \hlineB{2}
        \end{tabular}}
\end{table}

\begin{figure}[t!]
    \centering
    \includegraphics[width=\linewidth]{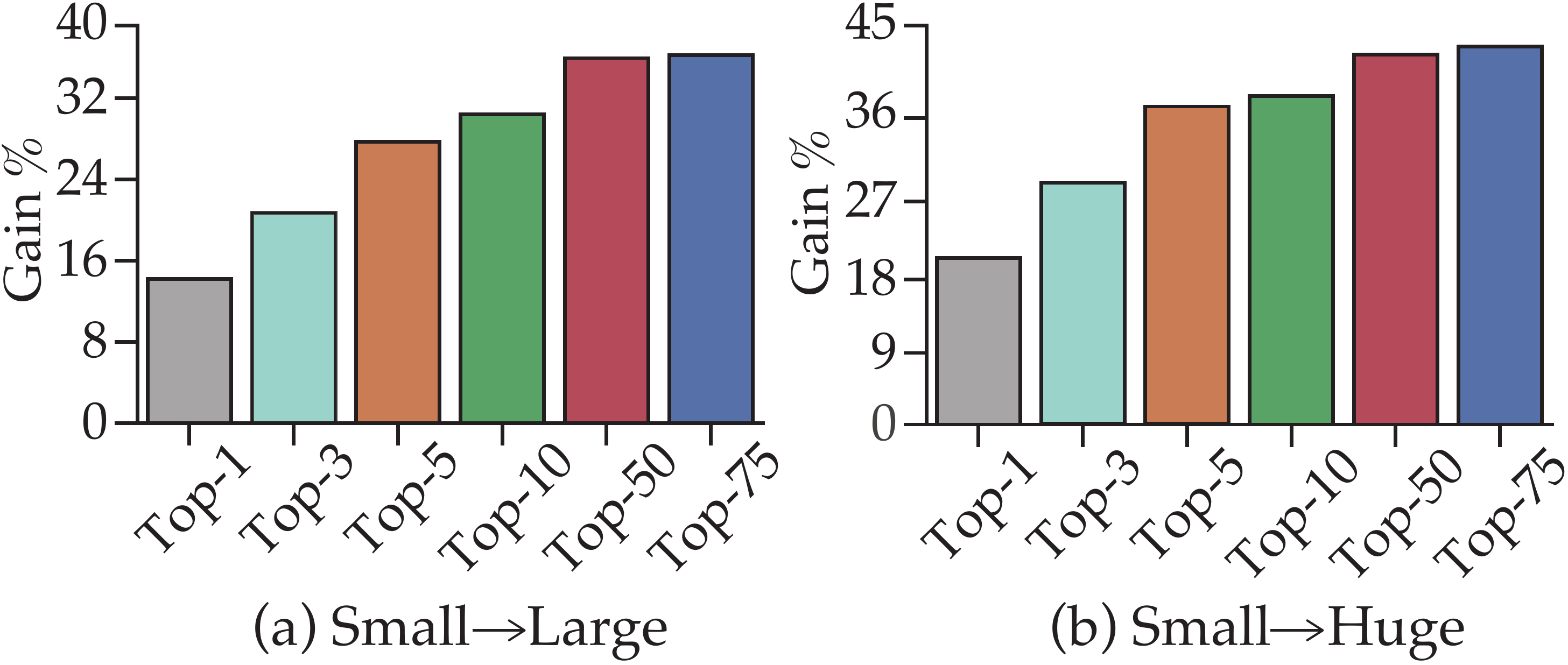}
    \caption{Performance gains over to default configuration for scale-up in Hadoop WordCount using top-1, top-3, top-5, top-10, top-25, top-50, and top-75 best configurations from the small workload. In both small$\rightarrow$large and small$\rightarrow$huge we notice even with top-1 we achieve a performance improvement over default. From top-3 to top-50 the performance gains increases. Beyond top-50 the improvements are marginal.}
    \label{fig:scale_up}%
\end{figure}%

\noindent\textbf{\textsf{\small RQ1-2.~How well does \tool \textit{Scale-Up} when exposed to much larger workloads?}}
We investigate this RQ in three steps: 
(i)~Evaluate the performance gain of Hadoop and Spark jobs for large and huge workloads when using the best configuration obtained for a small workload,  
{(ii)~Assess if any of the top-1 to top-75 configurations for a small workload achieve notable performance gains for large and huge workloads;} and  
(iii)~Perform a cost analysis to assess whether Scale-Up can save configuration exploration cost. 

\noindent
\textit{(i) Evaluating Performance Gain.}~%

Here, we determine whether good configurations found for \texttt{small} jobs provide comparable benefits for much larger jobs. Accordingly, with no additional sampling, we reuse the best configuration (\ie, Top-1) obtained for \texttt{small} workload (from \textit{RQ1-1}) on~\texttt{Large}~(10$\times$ larger) and~\texttt{Huge}~(100$\times$ larger) inputs for Hadoop jobs. For Spark, we use the best configurations (\ie, Top-1) obtained from \texttt{large} jobs to evaluate \texttt{Huge} inputs.

For Hadoop, in all five jobs, using the Top-1 best configuration, we see an average of 20.5\% and 19\% improvements under large and huge workloads respectively. We observe that, for WordCount, PageRank, and NutchIndex jobs, larger workloads exhibited better performance gains than the baseline improvements under the small workloads (see \tab{perf-improvement}), \eg, in Hadoop's WordCount, using \tool for small workloads (baseline) results in a performance gain of $12.5\%$ whereas small$\rightarrow$large results in a performance gain of $15.6\%$ while small$\rightarrow$huge results in a gain of $21.5\%$.

For Spark, using the Top-1 best configuration, we observed performance improvements for all five jobs ranging from $1.6\%$ to $16.8\%$. However, for {Sort} and {SVD}, there were limited improvements: $1.6\%$ and $1.9\%$ respectively. Spark jobs run very fast compared to Hadoop jobs. Although we have scaled up the workload 10 times, it seemed to be hard to achieve significant gains. Additional research is needed to better understand the scale-up potential of Spark jobs. Nevertheless, we saw an average performance improvement of 6.6\%, which we still believe is significant if it holds in practice at this scale.

\noindent%
\textit{(ii) Assessing variations in scale-up performance.}~
Larger workloads present several memory constraints in addition to other I/O overheads. Therefore, there is a potential for numerous variations in the scale-up performance (how well gains on small jobs scale up to similar gains on large jobs). To evaluate this, we assess how well each of the Top-1, Top-3, Top-5, Top-10, Top-50, and Top-75 best-performing configurations found in the small inputs performed on the much larger data sets. 

We observe that better performance gain is usually achieved for configurations that are among the Top-3, Top-5, Top-10, and Top-50 for small workloads as shown in \fig{scale_up}. We also observed that, beyond Top-50 configurations in the small workload (\eg, Top-75), there were no further improvements in scale-up. This was true for both Large and Huge workloads.

While it is true that running fifty production-scale jobs (instead of Top-1) as a final step of our approach would incur significant additional cost and compute resources, we find that doing so may produce a notable performance improvement (as illustrated in~\fig{scale_up}). This is a trade-off that depends on the available budget. If the budget is very low, industrial practitioners may use just to Top-1 best configuration from the small workload and still obtain improvements reported in~\tab{perf-improvement}. However, if additional budget is available, practitioners may choose to evaluate the Top-3 to Top-50 configurations to find larger improvements.

\begin{table}[bt!]
	\caption{{Average CPU time (secs) for default configuration in Hadoop for each of the studies job under three workload sizes. }}
	\label{tab:evaluation-time}	
\arrayrulecolor{black}
\setlength{\tabcolsep}{0.3cm}
	\centering
    \resizebox{0.9\linewidth}{!}{
	\begin{tabular}{@{}lcccc@{}}
        \hlineB{2}
	           & \textbf{Small} & \textbf{Large} & \textbf{Huge} & \textbf{\#EXP}\bigstrut\\ \hlineB{2}
	\textbf{WordCount}  & 166          & 862          & 9367        & \#3241 \bigstrut\\
	\textbf{Sort}       & 133          & 869          & 9891        & \#3318 \bigstrut\\
	\textbf{Terasort}   & 115          & 1056         & 8751        & \#2876 \bigstrut\\
	\textbf{Pagerank}   & 300          & 5657         & 13096       & \#3177 \bigstrut\\
	\textbf{NutchIndex} & 477          & 6596         & 11215       & \#4685 \bigstrut\\ 
	\hlineB{2}
	\end{tabular}}
\end{table}  
\begin{table}[t!]
    \centering
    \arrayrulecolor{black}
    \begin{center}
        \caption{Performance gains offered by \tool with the \textit{Scale-out} Hypothesis. 
            }
            \label{tab:scaleout}
        \begin{minipage}{\linewidth}
        \centering
        \textsc{\textbf{Hadoop}}\\[0.3em]
        \label{tab:hadoop-scaleout-mcmc}
        \resizebox{\linewidth}{!}{
        \begin{tabular}{lV{2}ccccc}
            \hlineB{2}
            \multicolumn{1}{cV{2}}{\multirow{2}{*}{\backslashbox{\textbf{Tgt.}}{\textbf{Src.}}}} & \multicolumn{4}{c}{\textbf{Similar Jobs}}  & \textbf{\begin{tabular}[c]{@{}l@{}}Diff.~Job\end{tabular}} \\ 
                & \textbf{WordCount} & \textbf{Sort} & \textbf{TeraSort} & \textbf{PageRank} & \textbf{Nutch} \\ \hlineB{2}
            \textbf{WordCount}        & \textit{\cellcolor{steel!08}{21.5\%}}      & 10.7\%        & \cellcolor{steel!08}{28.1\%} & 20.6\%        & 7.1\% \\ \clineB{2-2}{1} \clineB{4-4}{1}
            \textbf{Sort}      & 11.4\%      & \textit{15.8\%}        & 21.1\%      & 18.4\%        & 5.7\% \bigstrut\\ 
            \textbf{TeraSort}  & 10.4\%      & 1.8\%         & \textit{18.3\%}    & \cellcolor{steel!08}{29.9\%}  & 3.8\%  \bigstrut\\ \clineB{5-5}{1}
            \textbf{PageRank} & {20.8\%} & 23.8\%  & 23.4\%  & \textit{25.2\%} & \cellcolor{steel!08}{16.8\%} \bigstrut\\ \clineB{6-6}{1}
            \textbf{Nutch}     & 12.2\%      & \cellcolor{steel!08}{27.6\%} & 15.5\%        & 10.4\%     & \textit{14.2\%} \bigstrut\\ 
            \hlineB{2}
        \end{tabular}}
        \end{minipage}\\[0.5em]%
        \begin{minipage}{\linewidth}
        \centering
        \textsc{\textbf{Spark}}\\[0.5em]
        \label{tab:spark-scaleout}
        \resizebox{0.99\linewidth}{!}{
        \begin{tabular}{lV{2}ccccc}
            \hlineB{2}
                \multicolumn{1}{cV{2}}{\multirow{2}{*}{\backslashbox{\textbf{Target}}{\textbf{Source}}}} & \multicolumn{5}{c}{\textbf{Similar Jobs}} \\
                \textbf{}  & \textbf{WordCount}  & \textbf{Sort} & \textbf{TeraSort} & \textbf{RF}   & \textbf{SVD} \\ \hlineB{2}
                \textbf{WordCount}  & \textit{\cellcolor{steel!08}{5.8\%}}  & \cellcolor{steel!08}{50.8\%} & \cellcolor{steel!08}{22.8\%} & 5.9\%  & 1.8\%  \bigstrut\\ \clineB{2-4}{1}
                \textbf{Sort}     & 3.5\% & \textit{1.6\%} & 22.3\%  & 9.9\%  & \cellcolor{steel!08}{4.7\%}  \bigstrut\\ \clineB{6-6}{1}
                \textbf{TeraSort} & 5.1\% & 17.8\% & \textit{16.7\%}  & 9.9\%  & 3.1\%   \bigstrut\\ 
                \textbf{RF}       & 4.3\% & 20.1\% & 10.0\%   & \textit{7.2\%}  & 2.5\%   \bigstrut\\ 
                \textbf{SVD}      & 2.2\% & 23.8\% & 13.4\% & \cellcolor{steel!08}{23.4\%} & \textit{1.9\%} \bigstrut\\\hlineB{2}
        \end{tabular}}
        \end{minipage} 
        \end{center}
\end{table}%

\smallskip
\noindent
\textit{(iii) Cost Analysis.} 
The first three columns in~\Cref{tab:evaluation-time} show the total execution time (in seconds) across the master-slave nodes for each Hadoop job while our test-bed is configured with default configurations. 
For example,~\textit{WordCount} under a small workload takes about $33$ seconds per cluster (166.2/5, where 5 is the number of nodes of our cluster). However, it takes around $1873$ seconds with ``huge'' data, with the time difference of $1840$ seconds. The last column shows the number of dynamic evaluations of sampled configurations before \tool achieves the best configuration. Thus,  for \textit{WordCount} job, our scale-up strategy saves $1656$  hours ($(1873-33)*3241$ seconds) to find a better configuration using scale-up strategy. 
In total, for all the five jobs, the scale-up strategy saves about $9,600$ hours or $39.6$ times. In monetary terms, that amounts to about \$12,480 on the AWS EMR service with m4.xlarge EC2 instances. 

\begin{result}
    Configurations found using small workloads as proxies for sampling production-scale performance do tend to produce significant performance improvements (from 10\% to 20\%, on average) for much larger workloads and thus, save a significant amount of exploration cost.
\end{result}

\begin{table*}[!t]
    \arrayrulecolor{black}
    \setlength\abovecaptionskip{0.5\baselineskip}
    \setlength\belowcaptionskip{0.5\baselineskip}
    \setlength{\textfloatsep}{5pt plus 2pt minus 2pt}    
    \setlength{\tabcolsep}{1pt}
\scriptsize
\begin{center}
{
\caption{Most Influential Parameters for Hadoop Jobs. 
The numbers in the top row indicate the total performance gain achieved by the corresponding job. 
The numbers in the bottom row represent the percentage performance gain achieved by the corresponding parameters.}
\label{tab:sensitivity}
\resizebox{\linewidth}{!}{
\begin{tabular}{@{}c|c|c|c|c@{}}
\hline
\textbf{WordCount} (22.34\%) & \textbf{Sort} (19.77\%) & \textbf{TeraSort} (18.45\%) & \textbf{PageRank} (19.56\%) & \textbf{NutchIndex} (19.04\%) \\ \hline
\begin{tabular}[c]{@{}l@{}}memory.mb: 12.61\%\\ sort.spill.percent: 3.76\%\\ input.buffepercent: -0.48\%\\ job.max.split.locations: -0.67\%\\ yarn.app.aresource.mb: -1.54\%\end{tabular} &
\begin{tabular}[c]{@{}l@{}}input.buffepercent: 12.13\%\\ memory.mb: 4.29\%\\ io.seqfile.compress.blocksize: 1.83\%\\ io.file.buffesize: 1.58\%\\ java.opts: 1.11\%\end{tabular} &  
\begin{tabular}[c]{@{}l@{}}map.java.opts: 8.65\%\\ input.buffepercent: 7.22\%\\ task.io.sort.mb: 2.84\%\\ memory.mb: 1.75\%\\ io.seqfile.compress.blocksize: 1.58\%\end{tabular} & 
\begin{tabular}[c]{@{}l@{}}memory.mb: 10.82\%\\ input.buffepercent: 5.38\%\\ java.opts: 3.10\%\\ task.io.sort.mb: 2.38\%\\ memory.mb: 1.95\%\end{tabular} & 
\begin{tabular}[c]{@{}l@{}}map.java.opts: 9.44\%\\ task.io.sort.mb: 6.19\%\\ reduce.memory.mb: 4.92\% \\ map.memory.mb: 1.83\%\\ yarn.Rscheduleclass: 0.39\%\end{tabular}\\ \hline
\end{tabular}
}}
\end{center}
\textsuperscript{Note: Due to the page limit, we only list five most influential parameters.}

\end{table*}

\textbf{\textsf{\small RQ1-3.~After how many optimized runs of a job does \tool offer break-even benefits?}}
Previously, we showed that scale-up with both top-1 and top-50 configurations produces notable performance gains over default configurations.
However, such gains come with some initial exploration cost. For instance, to find the optimal configuration, \tool had to run a job a few iterations as per the allowed time budget, which is around 250 iterations with a small workload for word count job.  Further, for top-50 setting, we need to run an additional 50 production scale jobs with a large or a huge workload.
In this RQ, we investigate how many production scale jobs we need to run (with an unoptimized setting) to amortize this cost.




To answer this question, we use Hadoop’s WordCount as an example. First, we run \tool on a small workload for 12 hours (approx. 250 iterations). Second, we deploy WordCount for Large and with 3 different configuration settings: 
\bi[wide=0pt]
\item \textit{Configuration-1 (Baseline)}: Here we use the default configuration. We do not deploy \tool, therefore, this job incurs no initial overhead. 
\item \textit{Configuration-2 (Top-1)}: Here we use the best configuration obtained by running \tool on a small workload for 12 hours. This incurs an initial overhead of 12 hours.  
\item \textit{Configuration-3 (Top-50)}: Here we first run \tool for 12 hours to find the top-50 configurations on a small workload. Next, we run the top 50 configurations on large/huge workload to find the best configuration from among the top 50. Thus, in addition to the 12 hours taken to run \tool, this job incurs an additional initial overhead of having to run large/huge workloads 50 times. For large workload in hadoop WordCount, this adds 12 more hours of initial overhead. For Huge workload, this adds an initial overhead of 103 hours. 
\ei

\begin{figure}[t!]
\setlength\abovecaptionskip{\baselineskip}
\begin{center}
\includegraphics[width=\linewidth]{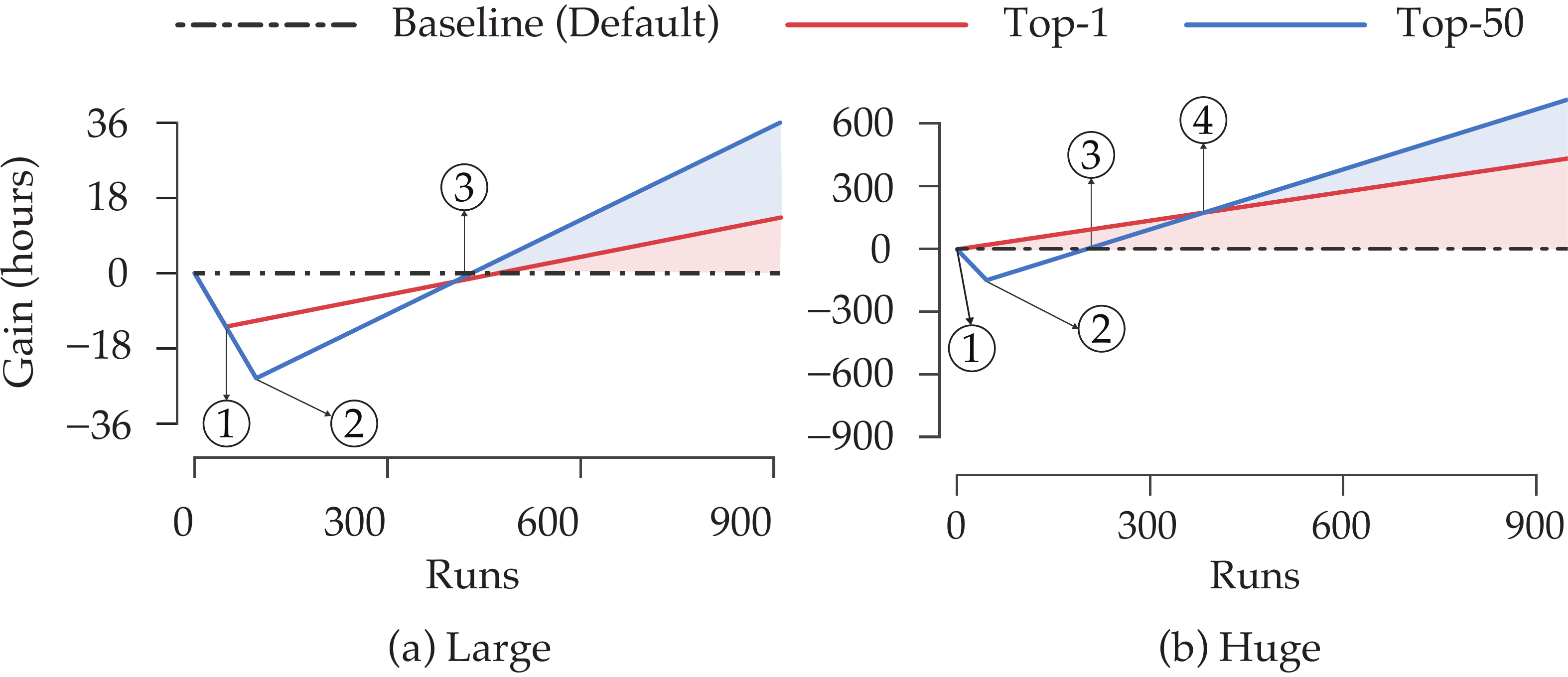}
\vspace{-2em}
\caption{Break-even analysis of scale out (top-1 and top-50) vs. default configuration. }
\label{fig:breakeven}
\end{center}
\end{figure}

\fig{breakeven} shows how many times we need to run a job in these experimental settings to amortize the initial exploration overhead with a trade-off curve. The x-axis shows job iterations in large/huge workload, and y-axis is the gain \wrt baseline. During the initial exploration phase, the gain will be negative because, during that time, one can run production-scale jobs. Once a better configuration is selected,  the job's performance starts gaining (reflected by positive slopes)  \wrt baseline. However, it takes some initial runs to amortize the overhead cost. 

The initial exploration to find top-1 configuration requires around 12 hours, which is our overhead. On average, a large workload takes 862 seconds to run. In 12 hours, we could have run the large workload around 48 times at the default configuration (see \circled{1} in~\fig{breakeven}(a)). To amortize the overhead at the top-1 setting, we need to run the large job additional 370 times. Thus, we  obtain a \textit{break-even} performance gain if the large workload were to be run more than 418 times (as shown by \circled{3} in~\fig{breakeven}(a)). In contrast, Top-50 incurs a total of 23 hours of initial overhead because we run \tool for 12 hours and then to run 50 configurations it takes another 11 hours after that to find the best configuration to use. This incurs a total cost equivalent to 98 runs of a large workload (shown by \circled{2} on~\fig{breakeven}(a)). In However, as noted in \fig{scale_up}, using top-50 offers an overall gain of 32\% as opposed to the 16\% offered by~top~1. Consequently, using top-50 after break-even offers more performance gain than top-1 (the region shaded \colorbox{steel!15}{\color{black} blue} is larger than baseline compared the region shaded by \colorbox{red!15}{\color{black}pink}). Thus, to amortize the overhead at the top-50 setting, we need to run the large job additional 306 times; we reach a break-even performance gain with default at 404 iterations. It is, therefore, generally better to expend the additional initial overhead to find the best configuration from among top-50.


\begin{table*}[t!]
\centering
\caption{\respto{3-1-B} Amortization costs for various workloads. The more compute intensive the workload is, the sooner we achieve the break-even point. Note: in each of the following setting, \tool was deployed for 12 hours.}
\label{tab:break-even}
\resizebox{0.8\textwidth}{!}{%
\begin{tabular}{r|r|r|r|r|r|r|r}
\hlineB{2}

\multicolumn{1}{c|}{\multirow{2}{*}{Job}} & 
\multicolumn{1}{c|}{Avg. Small Runtime} & 
\multicolumn{1}{c|}{\multirow{2}{*}{Workload}} & 
\multicolumn{1}{c|}{Default Runtime} & 
\multicolumn{1}{c|}{Gain} & 
\multicolumn{1}{c|}{\multirow{2}{*}{Setting}} & 
\multicolumn{1}{c|}{\multirow{2}{*}{Overhead}} & 
\multicolumn{1}{c}{Break-Even Point}\bigstrut\\

\multicolumn{1}{c|}{} & 
\multicolumn{1}{c|}{(seconds)} & 
\multicolumn{1}{c|}{} & 
\multicolumn{1}{c|}{(seconds)} & 
\multicolumn{1}{c|}{(\%)} & 
& 
\multicolumn{1}{c|}{} & 
\multicolumn{1}{c}{(wrt. Baseline)}\bigstrut\\ 

\hlineB{2}

\multirow{4}{*}{Word Count} & 
\multirow{4}{*}{166} & 
Large & 
897 & 
16 & 
Top-1 & 
48 & 
418\bigstrut\\

 &  
 & 
 Huge & 
 9367 & 
 21 & 
 Top-1 & 
 4 & 
 23\bigstrut\\

\cline{3-8}

 &  
 & 
 Large 
 & 
 897 & 
 32 & 
 Top-50 & 
 98 & 
 404\bigstrut\\
 
 &  
 & 
 Huge & 
 9367 & 
 41 & 
 Top-50 & 
 54 & 
 185\bigstrut\\\hline

\multirow{2}{*}{Sort} & \multirow{2}{*}{133} & Large & 869 & 7 & Top-1 & 49 & 749\bigstrut\\
 &  & Huge & 9891 & 15 & Top-1 & 4 & 30\bigstrut\\ \hline
\multirow{2}{*}{TeraSort} & \multirow{2}{*}{115} & Large & 1056 & 16 & Top-1 & 40 & 290\bigstrut\\
 &  & Huge & 8751 & 18 & Top-1 & 4 & 26\bigstrut\\ \hline
\multirow{2}{*}{NutchIndex} & \multirow{2}{*}{300} & Large & 6596 & 18 & Top-1 & 6 & 39\bigstrut\\
 &  & Huge & 11215 & 14 & Top-1 & 3 & 24\bigstrut\\ \hline
\multirow{2}{*}{PageRank} & \multirow{2}{*}{300} & Large & 5657 & 44 & Top-1 & 7 & 22\bigstrut\\
 &  & Huge & 13096 & 25 & Top-1 & 3 & 15\bigstrut\\ \hlineB{2}
\end{tabular}%
}
\end{table*}

For huge workloads, finding top-1 configuration with scale-up incurred an initial overhead of 12 hours. A huge workload, on average, takes 3 hours to run. Therefore, in 12 hours, we can deploy 4 instances of the huge workload with the default configuration before we can find a top-1 configuration with scale-up. Therefore, we achieve \textit{break-even} using top-1 almost immediately (after 4 runs) as shown by \circled{1} and the shaded \colorbox{red!15}{\color{black} pink} region in~\fig{breakeven}(b). On the other hand, Top-50 incurs a total of 12 hours of initial exploration overhead and then an additional 103 hours to run the top 50 configurations from scale-up to find the best configuration from among them to use we. This accounts for an equivalent of 54 runs of the huge job (shown by \circled{2} in \fig{breakeven}(b)). However, after as few as 250 runs, with using the configuration from top-50 we achieve break-even performance (see \circled{3} in \fig{breakeven}(b)). In huge, the performance gain obtained by using the best configuration from top-50 is 41\% (from \fig{scale_up}) compared to 19\% from top-1. This additional gain of using top-50 offers an overall break-even performance gain over top-1 if the huge job runs more than 380 times (see \circled{4} and region shaded by \colorbox{steel!15}{\color{black} blue} in \fig{breakeven}(b)).

\tab{break-even} highlights the break-even point for various workloads. Firstly, we notice that are compute intensive (NutchIndex and PageRank) achieve break-even \wrt using \tool much sooner. Specifically,  
\be
\item For Large workloads: It takes 39 and 22 default large runs before using \tool offers better amortized performance gain in NutchIndex and PageRank.
\item For Huge workloads: It takes 24 and 15 default large runs before using \tool offers better amortized performance gain in NutchIndex and PageRank.
\ee
Secondly, we observe that, in all cases, \tool offers break-even benefits much earlier for Huge workload. This is because, for huge workloads, the initial overhead of having to find a good configuration is far less detrimental than running unoptimized (default) configurations.

Typically a server (\eg AWS) is configured with \tool, where a job is expected to run numerous times. A web server, fo instance, is expected to process Millions of queries once it is configured. Hence, we hope the initial amortization cost is acceptable, given its benefits in the long run. Also, this cost is no more than the other learning-based approaches that need to collect samples to train a model. 

\begin{result}
\tool offers break-even performance gains over default configurations if a big data job runs more than 420 times (for large workloads). For workloads of size Huge, \tool offers break-even performance gain after only 4 runs. Overall, if a big data job runs more than 420 times (for a large workload) or 250 times (for a huge workload), then it is better to use the best configuration from top-50 even if the initial overhead is larger. 
\end{result}

\noindent\textbf{\textsf{\small RQ1-4.~How well does \tool \textit{Scale-Out} when exposed to different job types?}}

To further reduce exploration cost we assess if our scale-out hypothesis holds, \ie configuration found for one kind of job, A (e.g., Word Count), will also produce performance gains for a similar kind of job, B (e.g., Sort). We test this by evaluating performance gains for jobs of some job type, B, using configurations found for a job of some type, A, where the similarity between A and B is measured using~\Cref{sect:sim}. Among the five Hadoop jobs, we found \textit{WordCount}, \textit{Sort}, \textit{TeraSort} to be highly similar, that \textit{PageRank} is somewhat similar, and that \textit{NutchIndex} has low similarity with this group. ~\Cref{tab:scaleout} shows the results of the Scale-Out hypothesis.  

For example,~\Cref{tab:hadoop-scaleout-mcmc} shows that \tool found a configuration for \textit{WordCount (WC)} that improves its performance by $21.5\%$. When the same configuration is used for the similar target jobs: \textit{Sort}, \textit{TeraSort}, \textit{PageRank}, the performance gains achieved ($10.7\%$, $28.1\%$, and $20.6\%$ respectively) are close to the improvements found by their own best configurations. However, for \textit{NutchIndex}, which is not so similar, we see a performance gain of only  $7.1\%$, while it achieved 14.2\% gain while experimenting with its own best configuration. 
Similar conclusions can be drawn for Spark jobs (\Cref{tab:spark-scaleout}). 
A surprising outcome was that, in few cases, a better configuration found for one job, e.g., NutchIndex, yielded greater gains for another job, e.g., Sort (27.6\%) than the gain achieved by its improved configuration (15.8\% for Sort). 
We have left the analysis of such surprising behavior for future work.

\begin{result}
    Our scale-out hypothesis holds good, i.e., the configuration found with a representative job can bring significant performance gain for other similar jobs.
\end{result}

\noindent\textbf{\textsf{\small RQ1-5.~Which parameters have the highest impact on improving performance gain of \tool?}}
Here we study how sensitive performance gains are \wrt individual configuration parameters. From each best-found configuration, we set the value of each parameter back to its default value leaving all other improved parameter values unchanged and check to see how much the performance reverts to the baseline.  

For example, if ${perf}_\text{def}$ and $\mathit{perf}_\text{best}$ are the default and best performances (\ie CPU times) obtained by \tool for a job, then, the  performance improvement is 
$$\Delta_\text{best} = \left({\mathit{perf}_\text{def} - \mathit{perf}_\text{best}}\right)/{\mathit{perf}_\text{def}}$$ 
Next, to measure how sensitive the gain is \wrt a parameter $p_i$, we set $p_i$'s value back to default without changing the other parameter values from the best configurations. We measure the new performance \wrt to the default; Thus, $\Delta_\text{i} = (\mathit{perf}_\text{def} - \mathit{perf}_\text{i})/\mathit{perf}_\text{def}$. Then the sensitivity of parameter $p_i$ is the difference of performance improvement: 
$$\mathit{sensitivity}_i = \Delta_\text{best}  - \Delta_\text{i}$$
We conducted this analysis for all the parameters one by one for the Hadoop benchmark jobs using ``huge'' workloads. Table~\ref{tab:sensitivity} shows the results. The second row is the overall performance gain\footnote{We used a different cluster to do sensitivity analysis. So the overall performance could be slightly different from those in Table~\ref{tab:perf-improvement}.}.
The results suggest that performance improvement is sensitive to only a few parameters. However, no single parameter is responsible for most of the improvement. That is, 

\begin{result}
    The influences of individual parameters are limited and the overall improvements arise from the combinations of, or interactions between, multiple parameters in the configuration. 
\end{result}

These results suggest that, at least for Hadoop, higher-order interactions are present in the objective function and that these will need to be addressed by algorithms that seek high performing configurations.  Also, further improvements in sampling efficiency might be possible by focusing on a smaller subset of performance-critical parameters. However, we leave this to be explored in our future work.

\subsection*{RQ2.~{\rqd}}

Here we compare EMCMC with (i) random and (ii) genetic algorithm (GA) based evolutionary sampling strategies. A random approach samples a parameter value from the uniform distribution, \ie each value in the value range has the same probability to be selected. We have also implemented a GA based optimization strategy with the same cross-over and mutation strategies and the same fitness function as of EMCMC (See~\Cref{sect:emcmc}). For comparison, we run the baseline strategies to generate the same number of configurations and profile their performances with~\enquote{Small} data sets. We then conduct the scale-out validation to evaluate the performance gain in larger workloads.~\Cref{tab:hadoop-perf-gain} shows the performance gain of EMCMC over these two strategies. 

\begin{table}[!tbp]
    \setlength\abovecaptionskip{0.5\baselineskip}
    \setlength\belowcaptionskip{0.5\baselineskip}
\scriptsize
\begin{center}
  \caption{{Performance$^*$ gain of EMCMC over Genetic Algorithm (GA) and Random Sampling for Hadoop jobs}}
     \label{tab:hadoop-perf-gain}
     \arrayrulecolor{black}
     \setlength{\tabcolsep}{2.5pt}
    \resizebox{0.9\linewidth}{!}{ 
    \begin{tabular}{@{}lrrr|rrr@{}}
    \hline
          & \multicolumn{3}{c|}{Genetic Algorithm} & \multicolumn{3}{c}{Random} \bigstrut\\
          & {Small} & {Large} & {Huge} & {Small} & {Large} & {Huge} \bigstrut\\
    \hline
    WordCount & 92.31\% & 58.38\% & 84.61\% & 123.21\% & 61.66\% & 47.26\% \bigstrut\\
    Sort  & 2.12\% & -32.81\% & 53.75\% & 18.98\% & 44.53\% & 16.96\% \bigstrut\\
    TeraSort & 26.85\% & -6.96\% & 15.09\% & 136.21\% & 85.96\% & 64.72\% \bigstrut\\
    NutchIndex & 23.95\% & 62.26\% & 5.95\% & -31.83\% & 29.31\% & 18.96\% \bigstrut\\
    PageRank & -29.57\% & 155.80\% & 18.89\% & -0.67\% & 272.52\% & 77.75\% \bigstrut\\
    \hline
    Average & 0.63\% & 52.20\% & 31.16\% & 25.33\% & 105.34\% & 45.39\% \bigstrut\\
    \hline
    \end{tabular}}%
    \end{center}
    {$^*$Percentage performance improvement is computed as $\frac{\text{perf}_\mathit{emcmc} - \text{perf}_\mathit{ga/random}}{\text{perf}_\mathit{ga/random}}$}

\end{table}%

\begin{figure}[t!]
    \setlength\abovecaptionskip{\baselineskip}
    \begin{center}
    \includegraphics[width=0.48\textwidth]{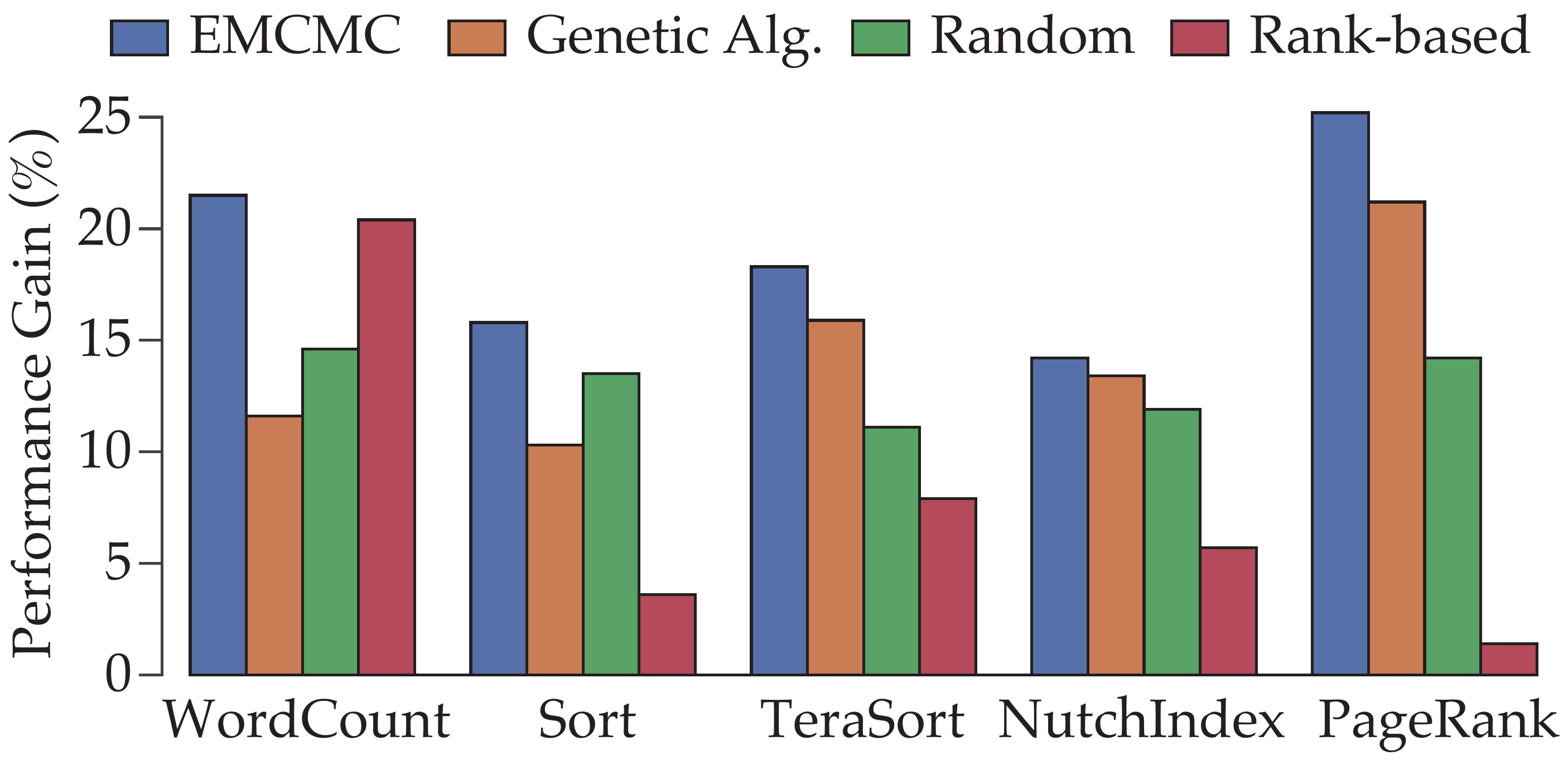}
    \caption{{EMCMC compares with other approaches in performance improvement for Hadoop Huge Workload.}}
    \label{fig:results}
    \end{center}
    \vspace{-.5cm}
\end{figure}

In general, for all the jobs, EMCMC based sampling performed better. 
For example, on average, EMCMC performed 52.20\% and 31.16\% better than GA for large and huge jobs. 
In comparison to random sampling, EMCMC performed $105.34\%$ and $45.39\%$ better, on average. 
~\Cref{fig:results} pictorially represents the results for \enquote{Huge} workload. 
The improvement of the performance of EMCMC over GA also gives us an estimate of how much the evolutionary part of EMCMC contributes to \tool's performance.

\begin{result}
  EMCMC based sampling strategy, on average, outperforms random (by up to 105\%) and genetic algorithm (by up to 52\%) based evolutionary sampling strategies to find better performing configurations for Hadoop.
\end{result}

\subsection*{RQ3.~\rqe}

To compare our approach with learning-based approaches, we choose the previous work of Nair \etal~\cite{nair-bad-learners} published in FSE 2017. We carefully choose this work as they also intended to find a near-optimal configuration rather than the best one, which is the most practical approach for a big-data job. 
They used a rank-based performance prediction model to find a better configuration. The authors argued that such a model works well when the cost of sampling is high, such as ours. They showed that compared to residual-based approaches, their model saves a few orders of magnitude of sampling cost and achieves similar and even better results.

In their experiments with larger systems having millions of configurations (\eg SQLite), the training pool covered 4500 configurations, including 4400 feature-wise and pair-wise sampled and extra 100 random configurations. We used the same approach\textemdash we randomly collected the same number of configurations as \tool to profile their performances (similar to RQ4) and used them as training. {We reused the model published by Nair \etal\footnote{\href{https://github.com/ai-se/Reimplement/tree/cleaned\_version}{https://github.com/ai-se/Reimplement/tree/cleaned\_version}}}. As they did, we ran each model $20$ times to find improved configurations.

For a fair comparison, following Nair et al., we evaluated both approaches by measuring rank difference (RD) between the predicted rank of a configuration and the rank of the training data (the profiled performance in our case). Table~\ref{tab:rank-description} shows the result. Here we ran each model $1000$ times to get enough data for the descriptive analysis. The results show that although the minimum RD is $0$, the average and maximum $RDs$ are $13.2$  and $408$ respectively, and the standard deviation is $24.4$. It means that this model could be largely wrong when trying to find high-performing configurations.

\begin{table}[t!]
\scriptsize
\centering
\arrayrulecolor{black}
\caption{{Descriptive rank differences of 1000 tests}}
\label{tab:rank-description}
\resizebox{0.6\linewidth}{!}{
\begin{tabular}{@{}lllll@{}}
\hline
\textbf{Job} & \textbf{Mean} & \textbf{Std} & \textbf{Min} & \textbf{Max} \bigstrut\\ \hline
\textbf{WordCount} & 13.2 & 24.4 & 0 & 408 \bigstrut\\ 
\textbf{Sort} & 28.7 & 42.6 & 0 & 391 \bigstrut\\ 
\textbf{TeraSort} & 14.3 & 19.1 & 0 & 171 \bigstrut\\ 
\textbf{NutchIndex} & 16.4 & 24.0 & 0 & 296 \bigstrut\\ 
\textbf{PageRank} & 9.5 & 16.7 & 0 & 158 \bigstrut\\ \hline
\end{tabular}}
\end{table}

None of the approaches we evaluated guarantee to find truly optimal configurations. So we discuss which approach can find the best candidate from all configurations checked. As we see from Table~\ref{tab:rank-description}, although a learning-based approach can find good configurations, it cannot guarantee the resulting one is the best. In some cases, the ranking mistake could be as large as $408$. On the other hand, our sampling-based approach can accurately find the best thanks to the dynamic evaluation and guided sampling strategy.

\noindent
\textbf{How much performance improvement one can gain by using Nair \etal's approach?} While our final goal is to improve system performance, we studied which approach can find better configurations, concerning how much performance one can gain. Suppose an engineer wants to use their approach to find a good configuration. She knows that all learning-based approaches have prediction errors. One possible way is to run such a model multiple times to rank configurations and then find the one with the best ranking on average across all tries. In this paper, we modified the tool released by Nair et al. to get the predicted ranking of configurations. We ran the above-described procedure $20$ times and find out the configuration with the highest rank in average. The last bar in Figure~\ref{fig:results} shows the performance improvement of the rank-based approach \wrt the default configuration. 
\tool performs $5.4\%$ to $1,700\%$ better than the ranked-based approach across five Hadoop jobs.

To understand why Nair \etal's approach doesn't perform well in finding good Hadoop configurations, we studied the accuracy of the trained models. In their implementation, the ranked-based model wraps a decision tree regressor as the under-hood performance prediction model. We checked the $R^2$ scores of these regressors, and it turns out that all scores are negative for all five jobs. It means that the trained model performs arbitrarily worse. 
This is not surprising because Hadoop's configuration space is complex, hierarchical, and high-dimensional; it is hard to learn a function approximating the objective function for such a space. A neural network-based regression model might work better. However, that would incur more sampling costs to gather adequate training samples.

\begin{result}
Compared to Nair et al's learning-based approach, our approach finds configurations with higher (from 5.4\% to 1,700\%) performance gains.
\end{result}

\section{Related Work}
\label{sect:related}

The related work broadly falls under two categories: (i) tuning big-data systems, and (ii) tuning traditional software. 

\noindent
\textbf{(i) Tuning big data framework.}

Starfish~\cite{herodotou2011starfish} is one of the initial works on Hadoop auto-tuning. It tunes parameters based on the predicted results of a cost model. However, the success of such a model depends largely on the underlying cost model, which is often hard to build. Liao \etal~\cite{liao2013gunther} have already proved that the predicted results of Starfish's cost model could vary largely under different task settings. They used a vanilla GA with only six important parameters to identify high-value configurations and beat~\cite{herodotou2011starfish}. We empirically showed EMCMC strategy performs better than a GA based approach. We further selected all parameters related to performance tuning, as without knowing how parameters interact, we cannot exclude any relevant parameter. Another line of work by  Babu et al.~\cite{babu2010towards} tune MapReduce parameters by sampling data from actual production, and thus, they optimize a job given a cluster and a fixed workload. Yu~\cite{yu2018datasize} optimize in-memory cluster over various data set sizes using hierarchical modeling and genetic algorithms. In contrast, \tool is more suitable to configure clusters where a diverse set of jobs are running under various sizes. Our assumptions are more generic and allow for learned configurations to be applied to a larger workload size (scale-up) and across similar jobs (scale-out).

A more recent line of research has explored multi-objective performance optimization focusing on performance goals such as throughput, latency, running time, etc. For example, Mishra \etal~\cite{mishra2015probabilistic}, propose LEO which uses an online probabilistic graphical model to learn a Pareto-optimal power/performance trade-off given a configuration. Compared to offline learning-based methods, LEO was demonstrated to perform better. More recently, Zhu \etal~\cite{zhu2017bestconfig} proposed BestConfig that uses divide-and-diverge sampling in conjunction with recursive bound and search to perform a multi-objective performance optimization with the help of a utility function to combine multiple objectives. \tool, on the other hand, is used to perform a search over a single objective and shows better performance gain (for runtime) compared to BestConfig on Hadoop's PageRank (the framework/job common to both the works).

\noindent
\textbf{(ii) Tuning Generic Software.} A large body of research exists on configuring generic software that uses different sampling+learning strategies. 
The main challenge to apply them directly to our case is the cost of dynamic profiling at the scale of big-data and the complexity of the configuration space. Here we systematically summarize these related work.

\textit{Configuration Sampling.} This step selects a subset of configurations based on different sampling strategies.  
For example, variations of  random~\cite{gogate2006new,guo2013variability,oh2017near-optimal} sampling are used to draw configurations uniformly from the configuration space. 
We have shown that \tool works much better than random sampling. 
Researchers also sampled test inputs targeting different regions~\cite{chakraborty2014distribution,chen2004adaptive}, or 
covering all the configurations satisfying some predefined constraints~\cite{johansen2012algorithm,lei2008ipog,marijan2013practical}. 
Kaltenecker \etal~\cite{kaltenecker2019distance} further proposes a distance-based sampling to sample the whole space uniformly.
Since big-data configuration space is quite complex and huge than previously studied systems, partitioning the configuration space is challenging and will require a significant amount of dynamic traces. Further, uniform sampling from different regions may not be necessary if the configurations that will lead to better performance is sparse. Instead, EMCMC based sampling strategy theoretically can approximate the global configuration space, and we showed that the guided approach can help to find a near-optimal configuration. Sampling-based approaches often select invalid configurations~\cite{apel2016feature}.  
To handle this problem, researches used constraint solvers to sample valid  configurations~\cite{henard2015combining,chakraborty2014distribution}.
Instead, we used an off-the-shelf configuration constraint checker~\cite{tang2017interpreted} (see~\Cref{sect:phthree}). 

\textit{Learning-based approaches.} 
A large body of previous works estimates system performance by learning prediction models~\cite{siegmund2012predictperf, guo2013variability, apel2013feature, nair-bad-learners,jamshidi2017transfer}. The accuracy of these models depends on the representativeness of training data. As shown in RQ5, for big-data systems, because of the complex high-dimensional configuration space, it is challenging to find a representative model. Also, collecting training samples is costly~\cite{weiss2008maximizing}. Existing logs from industrial uses of such systems are not necessarily useful as users tend to use the default, or at least very few, configuration settings~\cite{ren2013hadoop}.

Previous approaches also rely on the degree of parameter interactions. For example, Zhang et al.~\cite{zhang2015performance} assume all parameters are independent boolean variables and formulate the performance prediction problem as learning Fourier coefficients of a Boolean function. In contrast, Meinicke et al.~\cite{meinicke2016complexity} studied parameter interactions by running multiple configurations and comparing differences in control and data flow. They discovered that interactions are often less than expected but still complex enough to challenge search strategies. Siegmund et al.~\cite{siegmund2015performance} learned how parameter interactions influence system performance by adding interaction terms in learning models. This approach combines domain knowledge and sampling strategies to generate effective training data.  We have also seen evidence of parameter interactions in RQ2. However, unlike the predicting models, our search strategy is less affected by the parameter interactions as we have made no assumptions about such interactions. Thus, our work complements such previous efforts and present a novel search-based strategy for tuning big-data frameworks.

\noindent
\textbf{Other Applications.}
Many software engineering applications use sampling and optimization strategies in the past. For example, researchers used automated search to find valuable test cases~\cite{jia-higherorder-test,MacMinn-test-generation} and increase test coverage~\cite{carfast}.
    
Weimer et al.~\cite{weimer-icse2012} used genetic programming for program repair. 
Le~\cite{le2015-compiler-mcmc} used MCMC techniques to generate program variants with different control- and data-flows. 
Whittaker and Thomason~\cite{whittaker1994markov} used a Markov Chain model to generate test inputs to study software failure causes. 
Oh, et al.~\cite{oh2017near-optimal} worked to find good configurations for software product lines. 
Vizier~\cite{golovin2017google} was developed at Google for optimizing various systems like machine learning models.
Our work demonstrates the promise of similar approaches for the performance tuning of critical big-data systems.

There are some researches related to detecting software performance issues~\cite{forepost,main-effects-screening}. However, finding a better configuration for performance improvement and identifying performance issues in software are orthogonal problems. \c{T}{\u{a}}pu{\c{s}} \etal~\cite{active-harmony} propose a distributed tool to optimize a system’s resource utilization. We are interested in gaining higher performance given such resources. 

\section{Threats to validity}
\label{sect:threats}

\indent \textbf{Internal Validity.} Threats to the internal validity arise
from the experimental setup. First, 
the experimental results may be affected by uncontrolled factors on hardware platforms. In our experiments, we adopted some strategies to mitigate such unseen factors. For example, we make sure that no other programs are running while we are running experiments. We also choose a subset of all parameters to study with domain knowledge but we may have inadvertantly missed some important ones. To mitigate this threat, we referred to many previous works cited in this review paper~\cite{Bonifacio:research_review} on Hadoop, and have included all parameters studied by other researchers in our parameter set. It’s very expensive to run \tool many times because of the nature of ``big" data. Thus, running our algorithm enough times to get its statistic characteristics is an impossible mission. However, our two scale-up and scale-out hypotheses are created to mitigate this limitation. Although we didn’t conduct a statistical analysis, we tested our method on multiple jobs at different scales to demonstrate that our method works.

\textbf{External Validity.} We report results only for two big-data frameworks namely, Hadoop and Spark framework. Our results may not generalize to other frameworks. That said, Hadoop and Spark are among the most widely used big-data frameworks in HiBench, and we believe that the results are representative in other settings.

For scale-out we choose 10 representative jobs. The findings of this work may not apply to other job types. However, this paper chooses a diverse set of job types operating in different domains, \eg, \texttt{nutchindex} and \texttt{pagerank} are used in websearch; \texttt{svd} and \texttt{rf} are machine learning jobs; \texttt{sort} and \texttt{terasort} sort data; and finally \texttt{wordcount} operates on text data.

For the scale up hypothesis, our findings may vary for workloads that have significantly different memory consumption constraints and I/O overhead trade-offs. To address this threat 
we choose a diverse set of workloads with different memory consumption and I/O overheads. In particular, we choose a range of workload sizes that fit within our available hardware (Intel(R) Xeon(R) E5-2660 CPU, 32GB memory, and 2TB of SSD storage)\textemdash our smallest baseline workload (\texttt{small}) is 3MB and the largest workload (\texttt{huge}) is 3GB. Note that, the memory and I/O requirements for \texttt{small} and \texttt{huge} are vastly different. 
We also pick a diverse set of workloads. 
These diverse job types ensure that the scale-up may hold under different domains as well as under different workloads. 

Finally, due to the nature of dynamic evaluation, the experimental results may be affected by uncontrolled factors on hardware platforms. In our experiments, we adopted some strategies to mitigate such unseen factors. For example, we make sure that no other programs are running on the experimental platform while we are running experiments. We also run each dynamic evaluation three times to get average performance as a final result.

\textbf{Construct Validity.}%
~
{At various places in this paper, we made different engineering decisions, \eg, the range of values for each configuration parameter (from Phase-I in \tion{approach}), maximum number of generations ($\mathit{max_{gen}}$), etc. While these decisions were made using advice from the literature~\cite{Bonifacio:research_review} or based on engineering judgments, we acknowledge that other constructs might lead to other conclusions.}

\textbf{Evaluation Bias.}%
~
{In  RQ2 and  RQ3, we have shown the performance of \tool by comparing against Genetic Algorithms, Random Sampling, and Rank Based methods of Nair \etal to draw our conclusions. In choosing the comparisons, we chose those methods that (1) focus on single objective performance optimization, and (2) make available a replication package. The reported results hold for the software systems, evaluation metrics, and other performance optimization methods used for comparison in this paper. It is possible that with other Big-Data software systems and performance optimization methods methods, there may be slightly different conclusions. This is to be explored in future research.}

\section{Conclusions}
\label{sect:conclusion}
In this work, we proposed an EMCMC-based sampling strategy in combination with scale-up and scale-out tactics to cost-effectively find high-performing configurations for big-data infrastructures. We conducted and have reported results from carefully designed, comprehensive, and rigorously run experiments. The data that emerged provides  strong support for the hypothesis that our approach has strong potential to significantly and cost-effectively improve the performance of real-world big data systems. The data also strongly support the hypothesis that our approach outperforms several competing approaches. 

In this work, we had a single scalar objective function for each system: reducing CPU time for Hadoop and wall-clock time for Spark. However, in reality, there might be tradeoffs between performance improvements and other constraints (\eg cost). For example, user has to pay more money to Amazon EC2 for renting high performing systems. Whether techniques such as ours can be adapted to work in such situations remains a question for further study. 

\bibliographystyle{IEEEtran}
\balance
\bibliography{main}
\end{document}